# Thresholded Basis Pursuit: An LP Algorithm for Achieving Optimal Support Recovery for Sparse and Approximately Sparse Signals from Noisy Random Measurements. *

Venkatesh Saligrama   Manqi Zhao †
Department of Electrical and Computer Engineering
Boston University, MA 02215
Email: {srv, mqzhao}@bu.edu

**Abstract**

In this paper we present a linear programming solution for sign pattern recovery of a sparse signal from noisy random projections of the signal. We consider two types of noise models, input noise, where noise enters before the random projection; and output noise, where noise enters after the random projection. Sign pattern recovery involves the estimation of sign pattern of a sparse signal. Our idea is to pretend that no noise exists and solve the noiseless $\ell_1$ problem, namely, $\min \|\beta\|_1$ s.t. $y = G\beta$ and quantizing the resulting solution. We show that the quantized solution perfectly reconstructs the sign pattern of a sufficiently sparse signal. Specifically, we show that the sign pattern of an arbitrary k-sparse, n-dimensional signal $x$ can be recovered with $SNR = \Omega(\log n)$ and measurements scaling as $m = \Omega(k \log n/k)$ for all sparsity levels $k$ satisfying $0 < k \leq \alpha n$, where $\alpha$ is a sufficiently small positive constant. Surprisingly, this bound matches the optimal *Max-Likelihood* performance bounds in terms of $SNR$, required number of measurements, and admissible sparsity level in an order-wise sense. In contrast to our results, previous results based on LASSO and Max-Correlation techniques either assume significantly larger $SNR$, sublinear sparsity levels or restrictive assumptions on signal sets. Our proof technique is based on noisy perturbation of the noiseless $\ell_1$ problem, in that, we estimate the maximum admissible noise level before sign pattern recovery fails.

## 1 Introduction

The problem of recovering a sparse signal from noisy projections arises in many real world sensing applications [1, 2]. Motivated by these reasons we consider the problem of estimating $x$ based on noisy random projections, which we refer to as the *Output Noise Model*:

$$y = Gx + e \tag{1}$$

Here $x \in \mathbb{R}^n$ is a sparse signal with support size $k$. We assume that the minimum absolute value of the non-zero components of the sparse signal $x$ is bounded from below[1] by $x_{\min}$, which we assume without loss of generality to be equal to one. $G \in \mathbb{R}^{m \times n}$ is a matrix chosen from an IID Gaussian

---

*This research was supported by the ONR Young Investigator Program and Presidential Early Career Award (PECASE) N00014-02-100362, NSF CAREER award ECS 0449194, and NSF Grant CCF 0430983 and CNS-0435353
†Author names appear in alphabetical order. This paper was submitted to IEEE Trans. on Information Theory in March 2009.

[1]It is impossible to identify the support of a signal $x$ with arbitrarily small components from noisy measurements

ensemble with its components $G_{ij} \stackrel{d}{\sim} \mathcal{N}(0, \frac{1}{m})$. The noise vector $e$ is assumed to be Gaussian with IID components, with each component $e_j \stackrel{d}{\sim} \mathcal{N}(0, \frac{1}{SNR})$, and independent of $G$. Note that the term $SNR$ viewed in this normalized setting is also the inverse of the noise variance.

Notice that the setup of Eq. 1 parameterizes the problem in terms of two parameters namely, $SNR$ and the number of measurements.

In many cases such as system identification, active sensing, and sensor networks, noise can also arise at the input. We refer to such situations as the *Input Noise Model*. Motivated by these instances we also study recovery of $x$ from the following measurements:

$$y = Gz = G(x + w) \qquad (2)$$

where, $G$, $x$ are as in Equation 1. We let $w$ be an arbitrary deterministic $\ell_\infty$ bounded perturbation to a sparse signal $x$. $SNR$ in this case is the inverse of the square of the $\ell_\infty$ norm of the noise, i.e., $SNR = (\|w\|_\infty)^{-2}$. Note that in this setting, $x$ is the sparse approximation to the composite signal $z = x + w$. This situation is related to the so called approximately sparse or compressible signals (see [3]). The deterministic $w$ readily generalizes to the case when $w$ is a Gaussian random vector and we also state results for this case.

In *Compressive Sensing* the goal is to reconstruct the signal, $x$, with significantly fewer measurements $m = \dim(y)$ than the dimension of the signal $n = \dim(x)$, by exploiting signal sparsity. The noiseless problem ($e = 0$, $w = 0$) as well as its noisy counterpart have been the subject of intense research [4, 5, 2, 6, 7]. For the noiseless problem it is well known that if $x$ has fewer than $k$ non-zero components, it can be perfectly recovered if and only if every sub-matrix of $G$ formed by choosing $2k$ arbitrary columns of $G$ has full column rank. The reconstruction of an arbitrary sparse signal, $x$, from random projections, $Gx$ can be stated as a combinatorial optimization problem, which is known to be NP-hard [7]. In [4, 5, 2, 6, 8, 9] it is shown that for sufficiently small $k$, the so called $\ell_1$ relaxation can lead to exact recovery if the sensing matrix $G$ satisfies additional properties. For example, Donoho et. al. [4] show that $m = \Theta(k \log(n/k))$ measurements is sufficient for the recovery of any $k$ sparse vector of length $n$ provided the measurements are exact.

In the noisy case perfect recovery is generally impossible for continuous valued signals and an estimate $\hat{x}$ that closely approximates $x$ in some distance measure is desired. The distances commonly considered include the $\ell_2$ distance [3, 8], $\ell_1$ distance [10] and sign-pattern recovery [11, 12, 13, 14]. Sign-pattern recovery, which is the focus of this paper, deals with exactly recovering the sign pattern of the components of an arbitrary sparse signal $x$.

The problem of sign pattern recovery is motivated many problems such as the graph topology identification [12] where the mean squared error criterion provides an insufficient characterization of the solution. Two different approaches to the sign pattern recovery problem has been studied in the literature. One line of research has developed *algorithm-independent information theoretic performance bounds* to characterize fundamental limits on $SNR$, the number of measurements, and tolerable sparsity level required for exact sign pattern recovery from noisy measurements [15, 16, 13, 14]. In Aeron et. al [14] it is shown that for the setup considered in Equation 1 an $SNR = \Omega(\log(n))$ and number of measurements $m = \Omega(k \log(n/k))$ is both necessary and sufficient for sign pattern recovery of any signal $x$ with sparsity $k \leq \alpha n$, where, $\alpha$ is a sufficiently small positive number.

This paper adopts an algorithmic approach to sign pattern recovery and is based on our preliminary work [17]. Our algorithm consists of two steps: **(1)** In the first step we solve $\min \|\beta\|_1$ s.t. $y = G\beta$, with the data $y$ generated noisily, i.e., $y = Gx + e$ or $y = G(x + w)$. The solution to the optimization problem is then quantized. **(2)** In the second step we solve a least squares regression to improve our estimates obtained in the first step.

Remarkably, it turns out that our scheme essentially matches, in an order wise sense, the algorithm-independent necessary conditions on $SNR$ and the number of measurements required for sign pattern recovery. In comparison, as we describe in Section 1.2, our results are significantly stronger than bounds derived for other algorithmic approaches such as LASSO [12, 11] and Max-Correlation approach [13]. We also derive corresponding results for the input noise model based on our linear programming algorithm. While information theoretic results for this model has been developed [14], to the best of our knowledge, our paper is the first to report corresponding algorithmic results.

The paper is organized as follows. We first describe the notation used throughout the paper in Section 1.1. In Section 1.2 we present an overview of related work and also describe the main contributions of this paper. Section 2 describes the thresholded basis pursuit (TBP) algorithm. Section 3 develops the main results for sign pattern recovery for the input noise model. The proof for sign pattern recovery is broken up into several steps in Section 4. First, we establish sign pattern recovery in the linear sparsity regime. Then in the following section we develop sign pattern recovery for the general case where both linear and sublinear regimes are considered. In Section 5, these results are extended to the output noise model. Finally, we present some numerical results in Section 6.

## 1.1 Notation

We use capital letters to denote matrices and usually use small letters to denote signal vectors. The jth component of a signal $e$ is denoted as $e_j$. A matrix $U \in \mathbb{R}^{p \times q}$ where $p \geq q$ ($p \leq q$) is said to be orthonormal if each column (row) has unit norm and its columns (rows) are all orthogonal.

In the noisy sensing model $y = Gx + e$ or $y = G(x + w)$, the sensing matrix $G$ is of size $m \times n$. Correspondingly we have $x, w \in \mathbb{R}^n$ and $y, e \in \mathbb{R}^m$. We denote the support and sign pattern of $x$ as:

$$I_{supp} = \{j \mid x_j \neq 0\}; \quad I^+_{supp} = \{j : x_j > 0\}; \quad I^-_{supp} = \{j : x_j < 0\} \quad (3)$$

and denote

$$x_{\min} = \min_{j \in I_{supp}} |x_j| := 1 \quad (4)$$

as the minimum magnitude of $x$ on the support. We assume without loss of generality that $x_{\min} = 1$ and we use $x_{\min}$ or substitute the number one whenever convenient. The elements on the support is denoted by:

$$x_{supp} = (x_j)_{j \in I_{supp}} \quad (5)$$

The sparsity $k$ is the size of the support $|I_{supp}| := \#\{I_{supp}\}$. Whenever we consider a linear sparsity regime, namely, the sparsity, $k$, of the signal increases in proportion to the signal dimension, $n$, we introduce a parameter $\alpha$ to denote the sparsity ratio $\frac{k}{n}$. We also introduce a parameter $C$ to denote the dimension to measurement ratio $\frac{n}{m}$ whenever this ratio is a constant. Specifically, we let

$$\alpha = \frac{k}{n}, \ C = \frac{n}{m} \quad (6)$$

Let $\hat{x}$ be an estimate of $x$ based on $y$. We denote by:

$$\hat{I}_{supp} = \{j \mid \hat{x}_j \neq 0\}; \quad \hat{I}^+_{supp} = \{j : \hat{x}_j > 0\}; \quad \hat{I}^-_{supp} = \{j : \hat{x}_j < 0\} \quad (7)$$

We need the following notations to denote false alarms and misses:

$$N_m = |I_{supp}| - (|I^+_{supp} \cap \hat{I}^+_{supp}| + |I^-_{supp} \cap \hat{I}^-_{supp}|) \quad (8)$$

and
$$N_f = |\hat{I}_{supp}| - (|I^+_{supp} \cap \hat{I}^+_{supp}| + |I^-_{supp} \cap \hat{I}^-_{supp}|) \tag{9}$$

We also make a note of some probabilistic statements used in the paper. We use $\Pr(\cdot)$ to denote probabilities of events; $\mathbb{E}(\cdot)$ to denote expectations; $\mathbb{E}(z \mid v)$ to denote conditional expectation of $z$ given $v$; and $\mathbb{I}_\mathcal{V}$ to denote the indicator function for the set $\mathcal{V}$. We also often state that a random variable, $z$, satisfies

$$\|z\| \leq \gamma, \text{ w.p.} \geq \delta$$

to mean that, $\Pr(\|z\| \leq \gamma) \geq \delta$.

We adopt the family of Bachmann-Landau notations. Specifically, if we say $f(n) = \Theta(g(n))$ we mean that there is a positive integer $n_0$ and numbers $\lambda_1, \lambda_2 \in \mathbb{R}^+$ such that for all $n \geq n_0$ we have $\lambda_1 |g(n)| \leq |f(n)| \leq \lambda_2 |g(n)|$. By $f(n) = \Omega(g(n))$ we mean that there is a number $\lambda \in \mathbb{R}^+$ such that $\lim_{n \to \infty} \left| \frac{f(n)}{g(n)} \right| \geq \lambda$. By $f(n) = O(g(n))$ we mean that there is a number $\lambda \in \mathbb{R}^+$ such that $\lim_{n \to \infty} \left| \frac{f(n)}{g(n)} \right| \leq \lambda$. Finally, by $f(n) = o(g(n))$ we mean that $\lim_{n \to \infty} \left| \frac{f(n)}{g(n)} \right| = 0$.

## 1.2 Overview of Related Work & Our Main Contributions

The information-theoretic algorithm-independent necessary conditions for support recovery from noisy random projections for the setup described in Equation 1 have been developed by several authors [14, 15, 16, 13]. Sufficient conditions based on max-likelihood has appeared in [15, 14]. Specifically, the following result appears in Aeron, Saligrama & Zhao [14]:

**Theorem 1.1.** *No algorithm can recover the support for the model given by Equation 1 if $SNR = o(\log(n))$. Furthermore, if $SNR = O(\log(n))$ and if the number of measurements, $m = o(k \log(n/k))$, then support recovery is impossible. Conversely, if $m = \Omega(k \log(n/k))$ and $SNR = \Omega(\log(n))$ then the max-likelihood algorithm can exactly recover the support of the signal for the model given by Equation 1 with high probability for any sparsity $k \leq \alpha n$, where $\alpha$ is a sufficiently small positive number.*

The main contribution of this paper is that we can essentially achieve these bounds by basically using a linear programming algorithm. In recent years, researchers have focused on sign pattern recovery with convex relaxations such as LASSO [11, 12] as well as max-correlation based approach [13]. We will describe these approaches in some detail here.

**LASSO:** The $\ell_1$-constrained quadratic programming, commonly referred to as LASSO (*Least absolute shrinkage and selection operator*), solves the following optimization problem:

$$\min_{\beta \in \mathbf{R}^n} \frac{1}{2} \|y - G\beta\|_2^2 + \lambda \|\beta\|_1$$

where, $\lambda$ is a tuning parameter, which is carefully selected to realize a meaningful solution. The performance analysis of sign pattern recovery for the output noise model of Equation 1 has been recently characterized by Candes et. al. [11] and Wainwright [12]. To simplify the analysis these authors seek a bound on the number of measurements and $SNR$ to exactly recover the support. This means that we need to determine whether there is a suitable choice of $\lambda$, $SNR$ and $m$ such that the solution $\hat{x}$ of LASSO satisfies:

$$\hat{I}^+_{supp} = I_{supp}, \ \hat{I}^-_{supp} = I_{supp} \tag{10}$$

Remarkably it turns out that for a suitable choice of $\lambda$, Equation 10 is satisfied. However, this choice of $\lambda$ turns out to be conservative relative to the achievable bounds of Theorem 1.1. For instance, [12] requires high SNR for support recovery. In particular the author shows that in the high $SNR$ limit, the number of measurements $m = \Theta(k \log(n-k))$ is both necessary and sufficient for accurate sign pattern recovery. Candes et. al. [11], gives tighter SNR bounds for exact support recovery. They show for the setup of Equation 1 if $SNR \geq 200 \log(n)$; the maximum sparsity ratio is sublinear, i.e., $k = O\left(\frac{n}{\log(n)}\right)$; the support of the signal $x$ is uniformly distributed over all possible choices of support; then if the number of measurements, $m = \Omega(k \log(n))$, exact sign pattern recovery is guaranteed with high probability.

In contrast our analysis does not impose restrictions on the signal set and we admit sparsity ratios in the linear regime. We establish that an $SNR = \Omega(\log(n))$ and $m = \Omega(k \log(n/k))$ measurements are sufficient for exact sign pattern recovery with high probability. In practice these differences appears to be even more significant on numerical examples as illustrated in the Figure 1. The primary reason for this discrepancy can be attributed to the difference in our approaches. Our approach seeks to assert that the true support set of $x$ is always contained (with high probability) in the support set of the thresholded LP solution. The residual non-zero components of $x$ are significantly small and can be thresholded out. In other words, we do not attempt to exactly recover the support. In addition to these advantages we also point out that unlike LASSO we do not employ any tuning parameter on the quadratic penalty term. Indeed, from the analysis of [12, 11] it appears that this tuning parameter must be chosen as a function of the noise level, which our algorithm does not require. We describe this aspect in more detail in Section 6.

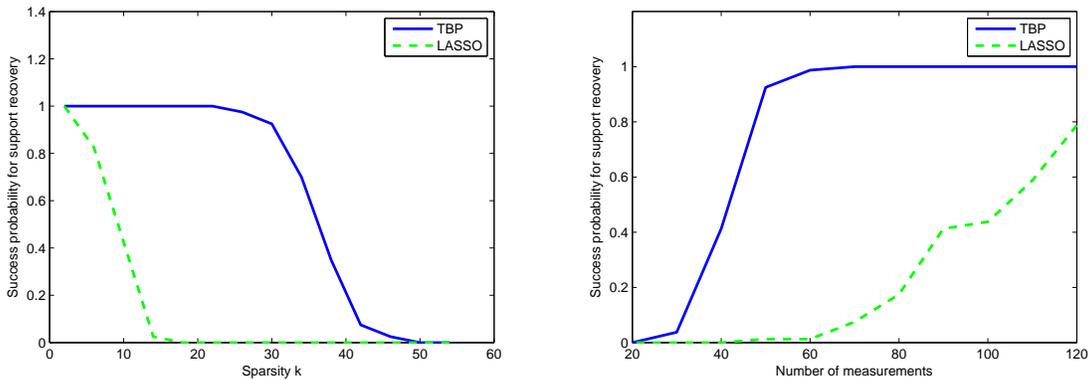

Figure 1: Left Figure: TBP vs. LASSO for perfect support recovery with different $k$ for the setup of Eq. 1 with $n = 200$ and $m = 100$ and $SNR = 6 \log(n)$. Sparsity $k$ is varied from 0 to 60. The success probability is computed based on 40 trials for each sparsity $k$. Success is declared if no false alarms or miss detections exist. Right Figure: $n$, $SNR$ fixed as in left figure and $k = 10$; $m$ is increased and the success probability is computed based on 80 trials.

**Max-Correlation Approach:** Fletcher et. al. [13], present necessary and sufficient conditions for sparsity pattern recovery based on maximum correlation estimator. The authors establish that the maximum correlation estimate is close to the necessary condition for sparsity pattern recovery. These results are stated under a different setup from that of Section 1.1. In particular, they introduce a different notion of $SNR$ and a notion of mean-to-average ratio (MAR). In particular, their sufficiency bound after appropriate substitutions turns out to be:

$$m > (8+\delta)\left(\frac{m}{x_{\min}^2} + \left\|\frac{x}{x_{\min}}\right\|_2^2\right)\log(n-k).$$

When this is translated to our setup with $x_{\min} = 1$, it turns out that this inequality will never hold asymptotically. It implies that this sufficient bound is actually a hybrid bound on both $x_{\min}$, $m$ and sparsity $k$. To further illustrate the issue consider the situation when $x_{supp} = (x_{\min}, 2x_{\min}, \cdots, kx_{\min})$ their bound on the number of measurements says that the number of measurements must scale as $m = \Omega(k^3 \log n)$.

## 2  The Thresholded Basis Pursuit Algorithm

Here we propose a new LP based algorithm, namely Thresholded Basis Pursuit (TBP). The analysis of the algorithm highlights the fact that SNR level is an important aspect in addition to the number of measurements. As described earlier in Section 1.2 our analysis shows that solution to the LP has relatively small non-zero components that do not belong to the support set of the true signal $x$. In addition we show that the support of the true signal is always recovered with high probability. There is a difference between the proof technique of [11, 12] and ours. [11, 12] investigate conditions such that LASSO solution leads to $x_j = 0$ for $j \notin I_{supp}$. In contrast we seek solutions such that components outside the support set are relatively small. This relaxation helps us in bridging the order gap between LASSO and the linearly achievable sparsity through max-likelihood decoder.

The algorithm is composed of two steps:

**Step 1:**  Apply the Basis Pursuit

$$\text{minimize} \quad \|\beta\|_1 \quad \text{subject to} \quad y = G\beta$$

**Step 2:** Threshold the solution $\hat{x}$ of Step 1 if and only if it's small, i.e.,

$$Q(\hat{x}_j) = \begin{cases} 0 & \text{if } |\hat{x}_j| < \frac{1}{2}x_{\min} := \frac{1}{2}; \\ \hat{x}_j & \text{otherwise.} \end{cases}$$

where we have assumed without loss of generality that $x_{\min} = 1$. The above algorithm will be referred to as *TBP*. We list below the main steps involved in the analysis of TBP. We then refine these estimates using a least squares regression (see Section 3.3).

**(A) Linear Regime for Input Noise Models:** It turns that the input noise model described in Eq. 2 in the linear regime namely, $k = \Omega(\alpha n)$ is the simplest to analyze. We analyze this case by considering the small noise limit when exact recovery can be guaranteed using mean-squared error bounds. It is easy to show (see Theorem 3.1 and Corollary 3.2) that in the small noise limit the required number of measurements to guarantee exact recovery scales as $O(k \log(n/k)) = O(n \log(1/\alpha))$. Freezing the number of measurements at this level we increase the level of noise and characterize the limit when the estimated support set does not contain the true support set. It turns out (see Theorem 3.4(A) and Theorem 3.5(A)) that this limit is reached precisely when $\|w\|_\infty = O(1/\sqrt{\log(n)})$. These results can be directly extended to the case when $w$ is a Gaussian random vector and we describe this situation subsequently.

(B) **General Case for Input Noise Model:** We then extend the results (see Theorem 3.4(B) and Theorem 3.5) obtained for the linear regime to the general sparsity case where $k \leq \alpha n$. The general sparsity case contains both the linear and the sublinear regime and therefore requires additional constraints on the noise[2]. This is because when we only assume an $\ell_\infty$ bound on $w$ and the sparsity of $x$ is sublinear in $n$ the ratio of signal power to noise power can be vanishingly small, i.e., as small as $\Theta(\log n/n)$. In the linear case this ratio is no smaller than $\Theta(\log n)$.

(B) **Output Noise Model:** We then combine linear regime **(A)** and the general case **(B)** for the input noise model and derive results for the output noise model(see Theorem 5.1). The basic idea here is to convert the output noise, $e$, into an equivalent input noise, $w$. The main complication here is that, in doing so, some correlation between $G$ and the equivalent noise $w$ is introduced and we describe how this can be handled in our analysis.

## 3 Input Noise Case

In this section, for clarity of presentation, we state the main results based on the input noise model $y = G(x + w)$. Similar results will be extended to output noise model $y = Gx + e$ in Section 5.

We provide a brief outline of the proof by further describing each of the steps described in the previous section.

1. **Weak Support Recovery:** Here we show (see Section 3.1) that the solution to the linear program (Step 1) of the TBP satisfies:

$$\|\hat{x} - (x + w)\|_2 = O\left(\frac{\|w\|_1}{\sqrt{k}}\right). \tag{11}$$

   The result implies that for sufficiently small $w$ the support of $\hat{x}$ contains the support of $x$.

2. **Support Detection:** We now increase the noise level while keeping the number of measurements frozen at the noiseless level and characterize the level at which the support set (and the sign pattern) of the LP solution no longer contains the support of $x$. The probability of missed detection is described in Section 3.2 in Theorem 3.4. We then apply least squares regression (see Theorem 3.5) to ensure no false support detections. The proof of Theorem 3.4 appears in Section 4 and is based on the following steps:

   - **Null Space Characterization:** We observe that the LP minimization problem can be recast as follows(see Section 4.1):

$$\min_{v \in \mathbb{R}^{n-m}} \|x + w + Av\|_1$$

     where, the random matrix $A$ is in the null space of $G$ (recall components of $G$ are IID Gaussian distributed) and the rows of $A$ are normalized. Suppose $\hat{v}$ is the optimal solution. For the linear sparsity regime, namely, $k = \alpha n$ for $\alpha > 0$, it follows directly from the $\ell_2$ approximation error (Equation 11) that

$$\|A\hat{v}\|_2 = \|\hat{x} - (x + w)\|_2 = O(\sqrt{n})$$

     with high probability for a deterministic $w$ with $\|w\|_\infty = O(1)$. By exploiting the properties of the normalized matrix $A$ it turns out that $\|\hat{v}\|_2 = O(\sqrt{n})$.

---
[2]The pessimistic results for the input noise case is fundamental and has also been observed in Aeron et. al.[14]

- **Conditional Independence Lemma:** Note that if $A$ and $\hat{v}$ are independent then any component of the vector $A\hat{v}$ is $O(1)$. This would result in a manageable perturbation on the non-zero components of $x$. However, when $A$ and $v$ are correlated, which is the case here, some components of $A\hat{v}$ could be large. In Section 4.2 we show that $A$ and $\hat{v}$ are only weakly correlated and so, no component $Av$ can be large on the support set.

## 3.1 Weak Support Recovery Based on Mean-Squared Error Bound

Our main goal in this section is to: (a) Present a squared norm approximation result for the TBP algorithm; (b) Derive a weak support recovery result based on a squared norm distortion bound. By weak support recovery we mean that either a large fraction of the support can be recovered with $\|w\|_\infty = O(1/\sqrt{\log n})$ or the support can be completely recovered with sufficiently small noise, $w$.

We will need the so called restricted isometry property introduced in [6].

**Definition 3.1.** Given a matrix $G$ and any set of column indices $T$, we use $G_T$ to denote the $m \times |T|$ submatrix of $G$ composed of the corresponding columns in $T$. We further denote $x_T$ as the vector whose support is on $T$. Then we say that a matrix $G$ satisfies the Restricted Isometry Property(RIP) condition with parameter $\delta_k$ if

$$(1 - \delta_k)\|x_T\|_2^2 \leq \|G_T x_T\|_2^2 \leq (1 + \delta_k)\|x_T\|_2^2, \quad \forall\, x_T, \forall\, T \text{ s.t.} |T| \leq k \tag{12}$$

We can think of $\delta_k$ as a functional mapping that maps positive integers $k$ to the unit interval. Thus $\delta_{2k}$ refers to the RIP constant when the cardinality of the set $T$ is smaller than $2k$. Note that this definition only applies to column-normalized matrices.

The definition needs to be modified if the sensing matrix is not normalized. Throughout this paper, we assume $G$ is a column normalized Gaussian matrix, i.e., each entry $G_{ij}$ is i.i.d. sample drawn from $\mathcal{N}(0, \frac{1}{m})$. The RIP constants for this situation is described in [18]. For our purposes it turns out that we need $\delta_{2k} < 1/7$. For the Gaussian matrix assumed in this paper it turns out that,

$$\Pr(\delta_{2k} \leq 1/7) \geq 1 - \exp(-c_1 m), \text{ if } m \geq c_2(2k)\log\left(\frac{n}{2k}\right) \tag{13}$$

where $c_1, c_2$ are constants, which are independent of $n$, $m$, $k$.

**Theorem 3.1.** *Suppose the sparsity $k$ is such that $\delta_{2k} < \frac{1}{7}$. Consider the input noise model $y = G(x + w)$ with $x$ having sparsity $k$. Then the optimal solution $\hat{x}$ of the Basis Pursuit (i.e., the solution to Step 1 of TBP) satisfies the following inequality,*

$$\|\hat{x} - (x + w)\|_2 \leq C_s \cdot \frac{\|w\|_1}{\sqrt{k}}. \tag{14}$$

*where the constant $C_s$ depends only on $\delta_{2k}$. For $G$ drawn from an IID Gaussian ensemble as in Eq. 2 it follows that the above equation holds with probability greater than $1 - \exp(-c_1 m)$ for $m \geq c_2(2k)\log(n/2k)$.*

The proof techniques for this theorem are borrowed from [3] where the mean squared error bound of LASSO is derived. The detailed proof is in the appendix (Section 7).

*Remark* 3.1. Note that the requirement that $\delta_{2k} \leq \frac{1}{7}$ is stronger than the condition for noiseless case, $\delta_{2k} \leq \sqrt{2} - 1$, (see [19]). This is the cost we pay for noisy recovery.

We can translate the above result into characterizing the support error as follows.

**Corollary 3.2.** *Assume $w$ is $\ell_\infty$ bounded, i.e., $\|w\|_\infty \leq \epsilon x_{\min}$ for some $\epsilon < 1/2$ and $G$ is as in Eq. 2. Then the TBP algorithm ensures*

$$\max\{N_m, N_f\} \leq \frac{n}{\alpha}\left(\frac{2C_s\epsilon}{1-2\epsilon}\right)^2 \quad w.p. \geq 1 - e^{-c_1 m}$$

*where $N_m$ represents the number of miss-detected components and $N_f$ represents the number of false-alarms defined in Equation 8, 9.*

*Proof.* Without loss of generality we assume $x_{\min} = 1$ as described earlier. We only prove $N_m \leq \frac{n}{\alpha}\left(\frac{2C_s\epsilon}{1-2\epsilon}\right)^2$ and the bound of $N_f$ follows from the exact same reasoning. By definition of $N_m$, TBP algorithm misses $N_m$ components of $x$ as 0 in $I_{supp}$. This implies the $\ell_2$ error $\|\hat{x} - (x+w)\|_2$ is at least $\sqrt{N_m(\frac{1}{2} - \epsilon)^2}$ because if component $i$ is missed then $|\hat{x}_i - (x+w)_i|$ is at least $\frac{1}{2} - \epsilon$. Then by applying Theorem 3.1, we have,

$$\sqrt{N_m(\frac{1}{2} - \epsilon)^2} \leq C_s \frac{n\epsilon}{\sqrt{k}} \quad w.p. \geq 1 - e^{-c_1 m}.$$

Solving this inequality gives an upper bound of $N_m$. □

*Remark 3.2.* The above theorem provides an asymptotic bound to the miss-detection rate $\rho_m := \frac{N_m}{|I_{supp}|}$. Corollary 3.2 implies

$$\rho_m \leq \frac{1}{\alpha^2}\left(\frac{2C_s\epsilon}{1-2\epsilon}\right)^2.$$

We can see that $\epsilon$ controls the miss detection rate. For example, if $\epsilon = O(\frac{1}{\sqrt{\log n}})$ and $k$ scales with $n$, then $\rho_m \leq O(\frac{1}{\log n})$, which vanishes when $n \to \infty$. On the other hand, the condition $\epsilon = O(\frac{1}{\sqrt{\log n}})$ is just saying $SNR = \Omega(\log n)$. This implies that when $SNR = \Omega(\log n)$, the miss detection rate vanishes asymptotically via TBP.

Alternatively, perfect support recovery is achievable with sufficiently high $SNR$. The following result (Lemma 3.3) will be useful later.

**Lemma 3.3.** *Suppose $\|w\|_\infty \leq \epsilon$ and $\epsilon$ is sufficiently small. The $k$ columns of $G$ that correspond to the correct support are part of the optimal basis with probability $\geq 1 - e^{-c_1 m}$.*

*Proof.* We know from Theorem 3.1 that with probability $\geq 1 - e^{-c_1 m}$ Basis Pursuit ensures $\|\hat{x} - (x+w)\|_2 \leq C_s \cdot \|w\|_1/\sqrt{k}$. Now we choose a sufficiently small $\epsilon$ such that $\epsilon < \frac{1}{4}x_{\min}$ and $C_s \cdot n\epsilon/\sqrt{\alpha n/2} \leq \frac{1}{4}x_{\min}$. Given $\|w\|_\infty \leq \epsilon$, Theorem 3.1 implies

$$|\hat{x}_i - x_i| \leq |\hat{x}_i - (x_i + w_i)| + |w_i| \leq \|\hat{x} - (x+w)\|_2 + \epsilon \leq C_s \cdot \|w\|_1/\sqrt{n} + \epsilon \leq \frac{1}{2}x_{\min}.$$

Since Basis Pursuit is an LP algorithm, the optimal solution must be a vertex of the polytope. Denote $G_1$ as the optimal basis in $G$ for this optimal solution. Since the sign pattern of $x$ is correctly recovered with probability $\geq 1 - e^{-c_1 m}$ as shown above, the $k$ columns of $G$ that correspond to the correct support must be included in $G_1$ with this probability. Otherwise, if the $i$-th column ($i \in I_{supp}$) is not selected into the optimal basis, then $\hat{x}_i = 0$ but we know the correct $|x_i| \geq x_{\min}$, which contradicts the above inequality $|\hat{x}_i - x_i| \leq \frac{1}{2}x_{\min}$. □

## 3.2 Support Detection

The result in the previous section is asymptotic and does not provide conditions for exact support recovery. Quite surprisingly, we can prove this stronger result based on Lemma 3.3 and the theory of duality in linear programming. The precise setup of input noise model is defined as follows.

**Definition 3.2** (Input noise model). Sensing model $y = G(x + w)$ where $G$ is a Gaussian matrix with i.i.d. entries $\mathcal{N}(0, \frac{1}{m})$. We assume $w$ is a deterministic $\ell_\infty$ bounded noise $\|w\|_\infty \leq \epsilon_0$ where $\epsilon_0$ will be specified and always on the order $\Theta(1/\sqrt{\log n})$. Recall that we scale the signal $x$ such that $x_{\min} = 1$.

**Theorem 3.4.** *Consider the input noise model in Definition 3.2.*
*(**A**) Suppose the support size, $k$, of $x$ is in the interval $\alpha n/2 \leq k \leq \alpha n$ where $\alpha$ is a positive real number satisfying $\delta_{2\alpha n} < 1/7$ (see Eq. 12, 13). Also assume $\|w\|_\infty \leq \epsilon_0$ where*

$$\epsilon_0 := \frac{1}{2}\left(1 + \frac{2\sqrt{2}}{\alpha}\left(1 - \sqrt{\frac{C-1}{C}}\right)^{-1} C_s + \frac{2\sqrt{2}}{\sqrt{\alpha}}\left(1 - \sqrt{\frac{C-1}{C}}\right)^{-1}\sqrt{\frac{C}{C-1}} C_s \sqrt{2\log n}\right)^{-1}$$

*Then the TBP algorithm satisfies:*

$$\Pr(N_m = 0) \geq 1 - \left(\frac{1}{\sqrt{\pi \log n}} + 2.24^{-(n-m)} + 2e^{-c_1 m} + 2e^{-\frac{(\sqrt{n}-\sqrt{n-m})^2}{8}}\right).$$

*where the probability is taken with respect to the Gaussian IID ensemble $G$ described in Equation 2; and $c_1$ is described in Equation 13.*
*(**B**) Assume $\ell_\infty$ bound on noise $w$, namely, $\|w\|_\infty \leq \epsilon_0 := 1/(5C_s\sqrt{\log n})$. In this case we also impose the additional assumption $\|w\|_1 \leq k/\log n$, then for any arbitrary support size, $k \leq \alpha n$, there is constant $\tilde{C}$ such that for $m \geq \tilde{C} k \log(n/k)$ the TBP algorithm recovers the support with probability at least $1 - e^{-c_1 m} - \hat{C} e^{-c(n-m)} - e^{-\bar{c}m/e} - \frac{2}{\sqrt{\log n}}$ where $\hat{C}, \bar{C}, c, \bar{c}$ are constants.*

*Proof.* See Section 4. □

*Remark* 3.3. This theorem implies that the miss detection probability is exactly 0 w.h.p. but does not say anything about the number of false alarms. We leave the discussion on false alarms to the next subsection (see Theorem 3.5).

*Remark* 3.4. Note that the sublinear sparsity is not covered by the part (**A**) of the theorem. The reason can be attributed to relative increase in noise level. Note that when we only assume an $\ell_\infty$ bound on $w$ and the sparsity of $x$ is sublinear in $n$ the ratio of signal power to noise power can be vanishingly small, i.e., as small as $\Theta(\log n/n)$. This does not happen for the linear case and it is no smaller than $\Theta(\log n)$. For this reason we need to scale the noise power as well, which is the result of part (**B**). For the output noise model the scaling of noise power with the number of measurements occurs naturally as it will become clear in Section 5 and we do not need this constraint there.

## 3.3 Eliminating Non-Support Elements

Theorem 3.4 only ensures no miss-detection in the support. However, the number of false alarms can also be reduced to zero through a standard regression technique.

To achieve zero false alarms we take thrice the number of measurements required for support detection. We partition the measurements into two parts. The first $m$ measurements are used to estimate the support elements using TBP. Since the basic feasible solution[3] of a linear program can only have at most $m$ non-zero entries the support of $x$ can be identified to within $m$ elements. We next utilize $2m$ measurements in a regression problem to estimate the support of $x$ using a standard least squares algorithm. Our modified algorithm is as follows.

**Step 1:** We take $3m$ measurements where $m$ is the number of measurements used in Theorem 3.4. Next we partition the measurements into two parts, $y_1 \in \mathbb{R}^m$ and $y_2 \in \mathbb{R}^{2m}$ and also partition the sensing matrix correspondingly

$$G = \begin{bmatrix} G_1 \\ G_2 \end{bmatrix}$$

**Step 2:** We apply the TBP algorithm proposed with respect to the first $m$ measurements $y_1$. Denote $I$ as the indices of nonzero components from this step. The number of non-zero components is at most $m$ since the optimal solution to a linear program is a basic feasible solution.

**Step 3:** Using the second set of measurements $y_2$, we compute

$$\hat{x} = G_{2,I}^{\dagger} y_2 = G_{2,I}^{\dagger}(G_2(x + w_2))$$

where $G_{2,I}$ is the submatrix of $G_2$ that comprises the columns in index set $I$ and $G_{2,I}^{\dagger} = (G_{2,I}^T G_{2,I})^{-1} G_{2,I}^T$ represents the Moore-Penrose pseudo-inverse of $G_{2,I}$.

**Step 4:** We next threshold the solution $\hat{x}$ if its magnitude is small, i.e.,

$$Q(\hat{x}_j) = \begin{cases} 0 & \text{if } |\hat{x}_j| < \frac{x_{\min}}{2}; \\ \hat{x}_j & \text{otherwise.} \end{cases}$$

*Remark 3.5.* From Theorem 3.4, all the support components are included in $I$ w.h.p. after Step 2. The Steps 3, 4 are intended to eliminate those potential false alarms from $I$.

*Remark 3.6.* The simulation in Section 6 seems to suggest that this modified algorithm is unnecessary and TBP by itself is sufficient for both detecting the support and eliminating false alarms. However, our analysis requires this post-processing.

This modified algorithm (referred to as *TBP+OLS*) is guaranteed to exactly recover the sign pattern of signal $x$ w.h.p.

**Theorem 3.5. (A)** *Consider the setup of Theorem 3.4(A). The TBP+OLS algorithm described above results in zero false positives and negatives with probability:*

$$\Pr(N_m = 0, N_f = 0) > 1 - \frac{3}{\sqrt{\pi \log n}} - 2.24^{-(n-m)} - 2e^{-c_1 m} - 2e^{-\frac{(\sqrt{n} - \sqrt{n-m})^2}{8}} - e^{-\frac{(\sqrt{n} - \sqrt{m})^2}{8}}.$$

**(B)** *For the setup of Theorem 3.4(B). The TBP+OLS algorithm described above results in zero false positives and negatives with probability:*

$$\Pr(N_m = 0, N_f = 0) > 1 - e^{-c_1 m} - \hat{C} e^{-c(n-m)} - e^{-\bar{c} m/e} - \frac{4}{\sqrt{\log n}} - e^{-\frac{(\sqrt{n} - \sqrt{m})^2}{8}}.$$

*Proof.* See Appendix (Section 7). □

---

[3] Basic feasible solution in simplex method is a solution obtained by setting any $n - m$ variables to zeroes in a system of $m$ linear equations in $n$ variables, and solving for the values of remaining $m$ variables

# 4 Proof of Theorem 3.4: Sign Pattern Recovery for Input Noise

The proof can be broken down into three steps.
**(1)** For a sufficiently small noise level, $w$ we know that $I_{supp} \subset \hat{I}_{supp}$. As the noise level is increased this situation may continue to hold even when $\hat{I}_{supp}$ changes. However, we show in Lemma 4.1 that minimum noise level at which $\hat{I}_{supp}$ changes must result in $I_{supp} \not\subset \hat{I}_{supp}$, namely, we must first loose one or more of the support elements.
**(2)** We are then reduced to determining the minimum noise level before one or more of the support elements are lost. It turns out that this regime (noise levels before we loose support) is best characterized in an equivalent null space setting(see Section 4.1). The null space setting reveals a structural property of the optimal solution (see Lemma 4.4), namely, that the change in the estimated solution $\hat{x}$ as the noise level is increased satisfies certain conditional independence properties.
**(3)** The conditional independence property directly leads to computable bounds on the maximum perturbation, $|x_j - \hat{x}_j|$, $j \in I_{supp}$ as a function of noise level(see Section 4.2) for the linear regime. Extensions to the general case $k \leq \alpha n$ is then presented in the following section.

We establish the first step by considering unit vectors along a specific (but arbitrary) direction. To this end, consider a unit vector $w$ (in $\ell_\infty$ sense) and a scaling parameter $\epsilon$. We have,

$$y = G(x + \epsilon w), \ \|w\|_\infty = 1 \tag{15}$$

Note that this is the same model as in Equation 2 except that we have extracted the noise level into a separate variable $\epsilon$.

Let $G_T$ be the optimal basis associated with the optimal LP (i.e., Step 1 of TBP in Section 2) solution, where $T$ is the column index set of size $m$. Without loss of generality assume that the true support $I_{supp} = \{1, 2, \ldots, k\}$, i.e., the first $k$ components. Lemma 3.3 says that for sufficiently small $\epsilon > 0$ support detection is guaranteed with high probability, namely, $I_{supp} \subset T$. Fix a value $\epsilon$ and the vector $w$ for which support detection is guaranteed.

Note that for fixed $w$ the basis, $G_T \in \mathbb{R}^{m \times m}$, continues to be optimal for smaller values of $\epsilon$. Furthermore, as $\epsilon$ increases $G_T$ remains optimal until a column in $T$ violates the optimality condition. For convenience we denote by:

$$L = G_T^{-1} G = \begin{bmatrix} L_0 \\ L_1 \end{bmatrix} \tag{16}$$

where on the RHS we have partitioned $L$ into two submatrices, $L_0 \in \mathbb{R}^{k \times n}$ and $L_1 \in \mathbb{R}^{(m-k) \times n}$. Note that the optimal solution to Basis Pursuit has $m$ non-zero elements, which is $\hat{x}_T = x_T + \epsilon L w$ and $\hat{x}_{T^c} = 0$ for sufficiently small values of $\epsilon$. The perturbation on the support elements is given by:

$$\hat{x}_j = x_j + \epsilon(L_0 w)_j, \ j \in I_{supp}$$

Denote by $\gamma_0$ the following:

$$\gamma_0 = \min\{\epsilon : x_j + \epsilon(L_0 w)_j = 0, \text{ for some } j \in I_{supp}\} \tag{17}$$

**Lemma 4.1.** *Fix a vector $w$ and a scalar $\epsilon > 0$ such that support detection is guaranteed(which is guaranteed with high probability by Lemma 3.3). Denote the associated optimal basis by $G_T$. It follows that, $G_T$ remains optimal for all $\epsilon \in [0, \gamma_0]$.*

*Proof.* The proof follows from primal-dual characterization of optimality. Lemma 3.3 says that for sufficiently small $\varepsilon$, the above $\hat{x}$ ($\hat{x}_T = x_T + \varepsilon L w$ and $\hat{x}_{T^c} = 0$) is the optimal solution of Basis Pursuit (primal problem). We denote $\pi$ as the optimal solution of the following dual problem:

$$\text{maximize} \quad \pi^T y \quad \text{subject to} \quad -1 \leq \pi^T G \leq 1$$

Duality theory [20] says the optimal primal cost equals dual cost,

$$\pi^T G(x + \epsilon w) = \pi^T y = \|x_T + \epsilon Lw\|_1$$

We know from Lemma 3.3 that the reconstruction error $\epsilon Lw$ will not exceed $\frac{1}{2}x_{\min}$ for sufficiently small $\epsilon$. This implies for $i \in \{1, 2, \cdots, k\}$, the sign of $x_i + \varepsilon(Lw)_i$ is determined by $x_i$. Therefore we have,

$$\pi^T G(x + \epsilon w) = \|x_T + \epsilon Lw\|_1 = \sum_{i=1}^{k} sgn(x_i)(x + \epsilon Lw)_i + \sum_{i=k+1}^{n} |(Lw)_i|.$$

From *complementary slackness*, $(\pi^t G)_i = sgn(x_i)$ for any index $i$ in the support(i.e., $i = 1, 2, \cdots, k$). Then the above equation can be further simplified to

$$\pi^T G\epsilon w = \sum_{i=1}^{k} sgn(x_i)(\epsilon Lw)_i + \sum_{i=k+1}^{n} |(\epsilon Lw)_i|$$

We now consider any positive $\gamma < \gamma_0$. Multiplying by $\gamma/\epsilon$ and adding $\sum_{i=1}^{k} sgn(x_i)x_i$ to both sides, we will have

$$\pi^T G(x + \gamma w) = \sum_{i=1}^{k} sgn(x_i)(x + \gamma Lw)_i + \sum_{i=k+1}^{n} |(\gamma Lw)_i| \qquad (18)$$

By definition of $\gamma$ we know that, $(x + \gamma Lw)_i$ has the same sign as $x_i$ for $i \in I_{supp}$. Consequently, the RHS of Equation 18 is exactly $\|x_T + \gamma Lw\|_1$ and the whole equation can be rewritten as,

$$\pi^T y = \|x_T + \gamma Lw\|_1$$

which exactly implies the primal cost equals dual cost for the primal-dual pair $(x + \gamma Lw, \pi)$. Therefore, we do not switch the optimal basis when noise is scaled upto $\gamma$. Now since $\gamma < \gamma_0$ can be arbitrary the result holds for the limiting value $\gamma_0$ as well. □

The task remaining reduces to determining the gain of the operator $L_0$. To this end we pass into a null space characterization.

## 4.1 Null Space Characterization

We first quote a classical result for Grassmanian manifolds (see Theorem 2.2 of [21] for more details).

**Lemma 4.2.** *There is a unique distribution on m-dimensional subspaces of $\mathbb{R}^n$ that is invariant under orthogonal transformations. A subspace from this distribution can be generated as:*

1. *The range of a random orthonormal $n \times m$ matrix with the orthogonal invariant(OI) distribution;*

2. *The orthogonal complement of the range of a random orthonormal $n \times (n - m)$ matrix with the OI distribution;*

3. *The range of a standard Gaussian random $n \times m$ matrix; or*

4. *The null space of a standard Gaussian random $(n - m) \times n$ matrix.*

Lemma 4.2 provides the tool for converting the original problem formulation into a null space characterization.

**Lemma 4.3.** *Suppose $y = G(x+\varepsilon w)$. There is a one-to-one correspondence between the constrained optimization problem*

$$\min_{\beta} \|\beta\|_1 \ s.t. \ y = G\beta \tag{19}$$

*and the unconstrained optimization problem*

$$\min_{v} \ \|x + \varepsilon w + Av\|_1 \tag{20}$$

*such that the optimal solution $\hat{x}$ of Equation 19 and the optimal solution $\hat{v}$ of Equation 20 satisfy $\hat{x} = x + \varepsilon w + A\hat{v}$. Moreover, the entries of $A \in \mathbb{R}^{n \times (n-m)}$ can be regarded as i.i.d. Gaussian samples from $\mathcal{N}(0, \frac{1}{n-m})$ if the entries of $G$ is i.i.d. Gaussian samples from $\mathcal{N}(0, \frac{1}{m})$.*

*Proof.* Choose $A$ as the $n - m$ dimensional null space of $G$. Note that it follows from applying parts (3) and (4) of Lemma 4.3 that we can realize $A$ as an IID Gaussian matrix with the specified properties (also see [21]). Then any $\beta$ satisfying $y = G\beta$ can be written as $\beta = x + \varepsilon w + Av$, where $v$ is an $(n - m)$ dimensional vector. This implies the original LP algorithm

$$\min \|\beta\|_1 \ s.t. \ y = G\beta$$

can be converted into the following equivalent unconstrained optimization problem:

$$\min_{v} \ \|x + \epsilon w + Av\|_1 \tag{21}$$

More importantly, Lemma 4.2 implies that the entries of $A$ can be characterized as i.i.d Gaussian random variables. Finally, we note that the global normalization factor $\frac{1}{m}$ or $\frac{1}{n-m}$ on the Gaussian distribution will not influence the result of Lemma 4.2. □

For convenience we denote $\hat{v}$ as the optimal solution of the null space problem:

$$\hat{v} = \mathrm{argmin}_v \|x + \epsilon w + Av\|_1 \tag{22}$$

Based on the above characterization the value $\gamma_0$ of Equation 17 can be equivalently cast in the null space. First we note that

$$\epsilon w_j + \sum_{k=1}^{n-m} A_{jk}\hat{v}_j = \epsilon(L_0 w)_j, \ j \in I_{supp}$$

because both of them represent the reconstruction error in the support. Based on the equality we have

$$\|\epsilon L_0 w\|_\infty \leq \epsilon + \max_{l \in \{1,\cdots,k\}} |\sum_{k=1}^{n-m} A_{lk}\hat{v}_k| \tag{23}$$

This implies that we are left to understand how $|\sum_{k=1}^{n-m} A_{lk}\hat{v}_k|$ scales with increasing $\epsilon$. Our main result of this section characterizes a structural property of the optimal solution. It establishes weak dependence between optimal solution and the elements of the support set.

**Lemma 4.4.** *Assume $\hat{v}$ is the optimal solution to Equation 22 when $\epsilon \leq \gamma_0$, where $\gamma_0$ is defined in Equation 17. Denote $F = \sum_{i=1}^{k} sgn(x_i)A_i$ where $A_i$ represents the ith row of $A$. Then if the RIP condition is satisfied (i.e., $k \leq \alpha n$ and $\delta_{2\alpha n} \leq \frac{1}{7}$), the optimal solution $\hat{v}$ is only determined by $A_{k+1}, A_{k+2}, \cdots, A_n$ and $F$.*

*Proof.* We know from Lemma 4.1 and Lemma 3.3 that if RIP condition is satisfied the optimal $\hat{v}$ recovers the sign pattern of $x$ when $\epsilon \leq \gamma_0$ in Equation 22. This implies the sign of $x_i + \epsilon w_i + A_i\hat{v}(i = 1, \cdots, n)$ is $sgn(x_i)$. Now consider a small neighborhood $N(\hat{v})$ of $\hat{v}$ such that the sign of $x_i + \epsilon w_i + A_iv(i = 1, \cdots, n, v \in N(\hat{v}))$ does not change in this neighborhood. Then $\min_{v \in N(\hat{v})} \|x + \epsilon w + Av\|_1$ is equivalent to

$$\min_{v \in N(\hat{v})} \sum_{i=1}^{k} sgn(x_i)(x_i + \epsilon w_i + A_iv) + \sum_{i=k+1}^{n} |(\epsilon w + Av)_i| \quad (24)$$

For linear optimization, local optimum is also the global optimum. Therefore, by neglecting the constant term $\sum_{i=1}^{k} sgn(x_i)(x_i + \varepsilon w_i)$, $\hat{v}$ is also the optimal solution to

$$\min \sum_{i=1}^{k} sgn(x_i)(A_iv) + \sum_{i=k+1}^{n} |(\epsilon w + Av)_i| = \min \left\{ Fv + \sum_{i=k+1}^{n} |(\epsilon w + Av)_i| \right\} \quad (25)$$

where $F = \sum_{i=1}^{k} sgn(x_i)A_i$ is defined in the assumption of the lemma. This implies that the optimal solution $\hat{v}$ depends on $A_1, A_2, \cdots, A_k$ only through their sum $F$. In other words, $\hat{v}$ is only a function of $A_{k+1}, A_{k+2}, \cdots, A_n$ and $F$ as long as the RIP condition $\delta_{2\alpha n} \leq \frac{1}{7}$ is satisfied. $\square$

*Remark* 4.1. The above result implies that there is only a weak dependence between the optimal solution, $\hat{v}$ and the rows of $A$ corresponding to the support elements.

## 4.2 Linear Sparsity Case: Proof of Theorem 3.4(A)

In this Section we only deal with the case $\frac{1}{2}\alpha n \leq k \leq \alpha n$ based on the result of Lemma 4.9. Our task is to determine the maximum tolerable $\epsilon$ or alternatively, compute $\gamma_0$ (see Eq. 17). Our main result in this section is as follows:

**Theorem 4.5.** *Consider the linear sparsity case described above. Then $\gamma_0 = \epsilon_0$ where $\epsilon_0$ is specified in Theorem 3.4(A) which is on the order of $\Theta(\frac{1}{\sqrt{\log n}})$.*

We establish the result through a sequence of steps. First, we need the following standard result on singular values for Gaussian matrices.

**Lemma 4.6.** *Suppose $A \in \mathbb{R}^{n \times (n-m)}$ is a Gaussian matrix with i.i.d. entries drawn from $\mathcal{N}(0, \frac{1}{n-m})$. Suppose $n = Cm$ where constant $C$ satisfies $1 < C < \infty$, then we have*

$$\|Av\|_2 > \frac{1}{2}\left(\sqrt{\frac{C}{C-1}} - 1\right)\|v\|_2, \quad \text{for all } v \in \mathbb{R}^{n-m}$$

*with probability $\geq 1 - e^{-\frac{(\sqrt{n}-\sqrt{n-m})^2}{8}}$.*

*Proof.* This lemma is a direct Corollary of a result in [22]. In [22], it is proved that $\sigma_{\min}$ has the concentration property:

$$\mathbb{P}\left(\sigma_{\min} \geq \sqrt{\frac{C}{C-1}} - 1 - t/\sqrt{n-m}\right) \geq 1 - e^{-t^2/2}.$$

We set $t = \frac{1}{2}\left(\sqrt{n} - \sqrt{n-m}\right)$ and the lemma follows. $\square$

**Definition 4.1.** Assume $\hat{v}$ is the optimal solution of Equation 22. We denote by $\mathcal{A}_0$ the intersection of two sets:

$$\begin{aligned}
\mathcal{A}_0 &= \left\{A : \|A\hat{v}\|_2 \leq C_s \cdot \frac{\|\epsilon w\|_1}{\sqrt{k}}\right\} \\
&\cap \left\{A : \|Av\|_2 > \frac{1}{2}\left(\sqrt{C/(C-1)} - 1\right)\|v\|_2, \forall\, v \in \mathbb{R}^{n-m}\right\}
\end{aligned}$$

Intuitively speaking, $\mathcal{A}_0$ contains all those well-behaved matrices $A$ such that Basis pursuit results in good solution in $\ell_2$ sense and the smallest singular value is lower bounded. We have the following property for $\mathcal{A}_0$.

**Lemma 4.7.** *Assume $\epsilon < \gamma_0$ in Equation 22. Then we have*

1. *If $A \in \mathcal{A}_0$, then $\hat{v}$ is only determined by $\{F, A_{k+1}, \cdots, A_n\}$ where*

$$F = \sum_{i=1}^{k} sgn(x_i) A_i \tag{26}$$

*In other words $\hat{v}$ is conditionally independent of $A_l$ for $1 \leq l \leq k$ when conditioned on $F$.*

2. *The measure of $\mathcal{A}_0$ satisfies $\Pr(A \in \mathcal{A}_0) \geq 1 - e^{-c_1 m} - e^{-\frac{(\sqrt{n}-\sqrt{n-m})^2}{8}}$.*

*Proof.* The first part of the lemma follows directly from Lemma 4.4. For the second part, we note that if $G$ satisfies RIP condition and we convert the problem to null space characterization $\min_v \|x + w + Av\|_1$, the solution $\hat{v}$ will satisfy $\|A\hat{v}\|_2 \leq C_s \cdot \frac{\|w\|_1}{\sqrt{k}}$ via Theorem 3.1. This implies that $\Pr\{A : \|A\hat{v}\|_2 \leq C_s \cdot \frac{\|w\|_1}{\sqrt{k}}\} \geq 1 - e^{-c_1 m}$

Finally, we know from the concentration inequality in Lemma 4.6 that

$$\Pr\{A : \|Av\|_2 > \frac{1}{2}\left(\sqrt{C/(C-1)} - 1\right)\|v\|_2, \forall\, v \in \mathbb{R}^{n-m}\} \geq 1 - e^{-\frac{(\sqrt{n}-\sqrt{n-m})^2}{8}}$$

and hence the second part of the lemma follows. $\square$

We also need the tail probability of $\|F\|_2$.

**Lemma 4.8.** *Suppose $F = \sum_{i=1}^{k} sgn(x_i) A_i$, then we have,*

$$\Pr(\|F\|_2 \leq \sqrt{2k}) \geq 1 - 2.24^{-(n-m)}$$

*Proof.* We note that $\|F\|_2^2 = \sum_{j=1}^{n-m} \left(\sum_{i=1}^k sgn(x_i) A_{ij}\right)^2$. Then we can rewrite $\frac{n-m}{k}\|F\|_2^2$ as

$$\frac{n-m}{k}\|F\|_2^2 = \sum_{j=1}^{n-m} \left(\sum_{i=1}^k sgn(x_i) \sqrt{\frac{n-m}{k}} A_{ij}\right)^2.$$

Since each $A_{ij}$ is i.i.d Gaussian $\mathcal{N}(0, \frac{1}{n-m})$ from Lemma 4.2, $\left(\sum_{i=1}^k sgn(x_i) \sqrt{\frac{n-m}{k}} A_{ij}\right)$ is i.i.d standard Gaussian random variable. Therefore $\frac{n-m}{k}\|F\|_2^2$ is a $\chi^2$ distribution with $(n-m)$ degree of freedom. From the tail probability of $\chi^2$ distribution, we have

$$\Pr\left(\frac{n-m}{k}\|F\|_2^2 \leq 2(n-m)\right) \geq 1 - 2.24^{-(n-m)}.$$

□

**Lemma 4.9.** *Suppose, $A \in \mathcal{A}_0$ and we are in the linear sparsity regime, namely, $\frac{1}{2}\alpha n \leq k \leq \alpha n$ where $\alpha$ is an absolute constant such that $\delta_{2\alpha n} < \frac{1}{7}$. Then,*

$$\Pr\left(\max_{l \in \{1,\cdots,k\}} |A_l \hat{v}| > (d_1 + d_2\sqrt{2\log n})\epsilon\right) \leq \frac{1}{\sqrt{\pi \log n}} + 2.24^{-(n-m)} + 2e^{-c_1 m} + 2e^{-\frac{(\sqrt{n}-\sqrt{n-m})^2}{8}}$$

*where $\epsilon$ is the $\ell_\infty$ bound on input noise, $w$; $A_l$ is the lth row of $A$; and $(d_1, d_2)$ are absolute constants which only depend on $\alpha$ and $C$:*

$$d_1 = \frac{2\sqrt{2}}{\alpha}\left(\sqrt{\frac{C}{C-1}} - 1\right)^{-1} C_s, \quad d_2 = \frac{2\sqrt{2}}{\sqrt{\alpha}}\left(\sqrt{\frac{C}{C-1}} - 1\right)^{-1} \sqrt{\frac{C}{C-1}} C_s$$

*Proof.* It is not easy to directly bound $\Pr\left(\max_{l \in \{1,\cdots,k\}} |A_l \hat{v}| > (d_1 + d_2\sqrt{2\log n})\epsilon\right)$ because $A_l$ and $\hat{v}$ are (weakly) correlated in general. Therefore we introduce an auxiliary variable $\hat{v}^*$ as

$$\hat{v}^* = \arg\min_v \{Fv + \sum_{i=k+1}^n |(w+Av)_i|\} \tag{27}$$

Now $\hat{v}^*$ and $A_l(l \in \{1,,\cdots,k\})$ are independent given $F = f$. From Lemma 4.4 and Definition 4.1, $\hat{v} = \hat{v}^*$ if $A \in \mathcal{A}_0$. Moreover, if $A \in \mathcal{A}_0$, the $\ell_2$ norm of $\hat{v}^*$ can be bounded via applying Lemma 4.6 and Theorem 3.1:

$$\|\hat{v}^*\|_2 \leq 2\left(\sqrt{\frac{C}{C-1}} - 1\right)^{-1} \|A\hat{v}^*\|_2 \leq 2\left(\sqrt{\frac{C}{C-1}} - 1\right)^{-1} C_s \frac{n\epsilon}{\sqrt{\frac{1}{2}\alpha n}} \tag{28}$$

For simplicity of notation, we denote the RHS as $C'\sqrt{n}\epsilon$ by introducing $C' := 2\sqrt{2}\left(\sqrt{\frac{C}{C-1}} - 1\right)^{-1} C_s/\sqrt{\alpha}$.

Then, we can relate $\hat{v}$ and $\hat{v}^*$ from the law of total probability,

$$\Pr\left(\max_{l \in \{1,\cdots,k\}} |A_l \hat{v}| > (d_1 + d_2\sqrt{2\log n})\epsilon\right)$$
$$= \Pr\left(\max_{l \in \{1,\cdots,k\}} |A_l \hat{v}| > (d_1 + d_2\sqrt{2\log n})\epsilon, A \in \mathcal{A}_0\right) + \Pr\left(\max_{l \in \{1,\cdots,k\}} |A_l \hat{v}| > (d_1 + d_2\sqrt{2\log n})\epsilon, A \notin \mathcal{A}_0\right)$$
$$\leq \Pr\left(\max_{l \in \{1,\cdots,k\}} |A_l \hat{v}^*| > (d_1 + d_2\sqrt{2\log n})\epsilon\right) + \Pr\left(A \notin \mathcal{A}_0\right)$$

The second term is bounded by $e^{-c_1 m} + e^{-\frac{(\sqrt{n}-\sqrt{n-m})^2}{8}}$ and the remaining task is to bound the first term. To this end we fix an index $l \in \{1, 2, \ldots, k\}$ and define,

$$\mathcal{V}_1 = \{(A_l, \hat{v}^*) : |A_l \hat{v}^*| \geq (d_1 + d_2\sqrt{2\log n})\epsilon\}, \quad \mathcal{V}_2 = \{(\hat{v}^*, F) : \|\hat{v}^*\| \leq C'\sqrt{n}\epsilon, \|F\| \leq \sqrt{2k}\}$$

Let $\mathbb{I}_{\mathcal{V}_i}$, $i = 1, 2$ denote the corresponding indicator functions on these sets. Then, we can write

$$\begin{aligned}
\Pr\left(|A_l \hat{v}^*| > (d_1 + d_2\sqrt{2\log n})\epsilon\right) &= \mathbb{E}(\mathbb{I}_{\mathcal{V}_1}) = \mathbb{E}_{F,\hat{v}^*}(\mathbb{E}(\mathbb{I}_{\mathcal{V}_1} \mid F, \hat{v}^*)) \\
&= \mathbb{E}_{F,\hat{v}^*}(\mathbb{I}_{\mathcal{V}_2}\mathbb{E}(\mathbb{I}_{\mathcal{V}_1} \mid F, \hat{v}^*)) + \mathbb{E}_{F,\hat{v}^*}(\mathbb{I}_{\mathcal{V}_2^c}\mathbb{E}(\mathbb{I}_{\mathcal{V}_1} \mid F, \hat{v}^*)) \\
&\leq \mathbb{E}_{F,\hat{v}^*}(\mathbb{I}_{\mathcal{V}_2}\mathbb{E}(\mathbb{I}_{\mathcal{V}_1} \mid F)) + 2.24^{-(n-m)} + \Pr(A \notin \mathcal{A}_0) \quad (29)
\end{aligned}$$

where the first term in the last inequality follows from the Conditional Independence Lemma 4.7 and the second term in the last inequality follows from Lemma 4.8 and Equation 28. The first term in the above expression can be further simplified by noting that, $A_l \mid_F$ is a Gaussian random vector and $A_l \hat{v}^* \mid_F$ is a Gaussian random variable for any fixed $\hat{v}^*$. So, we compute the conditional means and variances:

$$\mathbb{E}(A_l | F = f) = sgn(x_l)\frac{f}{k}, \quad \Lambda_{A_l|f} = \frac{1}{n-m}\left(1 - \frac{1}{k}\right)I_{n-m} \quad (30)$$

Now for $(F, \hat{v}^*) \in \mathcal{V}_2$ we have $\|f\|_2 \leq \sqrt{2k}$ and $\|v\|_2 \leq C'\sqrt{n}\epsilon$. This leads to a bound on the conditional mean and variance of Gaussian variable $A_l \hat{v}^*|_{F=f}$. By applying the result in Equation 30 the absolute value of its mean is bounded by

$$|\mathbb{E}(A_l \hat{v}^* | F = f)| = |\frac{1}{k}f\hat{v}^*| \leq \frac{1}{k}\|f\|_2\|\hat{v}^*\|_2 \leq \frac{2}{k}\|f\|_2\left(\sqrt{\frac{C}{C-1}} - 1\right)^{-1} C_s\sqrt{\frac{2n}{\alpha}}\epsilon$$

where the first inequality follows from Cauchy-Schwartz inequality. Recall that $C$ is ratio $n/m$; $C_s$ is described in Equation 14; $\epsilon$ is the $\ell_\infty$ bound on the input noise, $w$.

Now by using the assumption $\frac{1}{2}\alpha n \leq k \leq \alpha n$ and $\|f\|_2 \leq \sqrt{2k}$, $|\mathbb{E}(A_l\hat{v}^*|F=f)|$ can be bounded by,

$$|\mathbb{E}(A_l\hat{v}^*|F=f)| = \frac{1}{k}\|f\|_2\left(\sqrt{\frac{C}{C-1}} - 1\right)^{-1} C_s\sqrt{\frac{2n}{\alpha}}\epsilon \leq \frac{2\sqrt{2}}{\alpha}\left(\sqrt{\frac{C}{C-1}} - 1\right)^{-1} C_s\epsilon = d_1\epsilon$$

And the variance is bounded through

$$\sigma^2_{A_l\hat{v}^*|F=f} = \frac{1}{n-m}\left(1 - \frac{1}{k}\right)\|\hat{v}^*\|_2^2 \leq \frac{4}{n-m}\left(\sqrt{\frac{C}{C-1}} - 1\right)^{-2} C_s^2 \frac{2n}{\alpha}\epsilon^2 = d_2^2\epsilon^2$$

Combining the above two bounds on mean and variance we can bound the first term in the final expression in Equation 29.

$$E(\mathbb{I}_{\mathcal{V}_1} \mid (F, \hat{v}^*)) = \Pr(|A_l\hat{v}^*| > d_1\epsilon + d_2\epsilon t | F = f, \hat{v}^* = v) \leq 2Q(t) \leq \frac{2}{t\sqrt{2\pi}}e^{-t^2/2}$$

when $\|f\|_2 \leq \sqrt{2k}$ and $\|v\|_2 \leq C'\sqrt{n}\epsilon$. Substitute $t$ with $\sqrt{2\log n}$ in the above equation and we have

$$\Pr\left(|A_l\hat{v}^*| > (d_1 + d_2\sqrt{2\log n})\epsilon | F = f, \hat{v}^* = v\right) < \frac{1}{k\sqrt{\pi \log n}} \quad \text{when } \|f\|_2 \leq \sqrt{2k}, \|v\|_2 \leq C'\sqrt{n}\epsilon.$$

Finally by applying a union bound when $\|f\|_2 \leq \sqrt{2k}, \|v\|_2 \leq C'\sqrt{n}\epsilon$ we get:

$$\Pr\left(\max_{l \in \{1,\cdots,k\}} |A_l \hat{v}^*| > (d_1 + d_2\sqrt{2\log n})\epsilon|_{F=f,\hat{v}^*=v}\right) \leq k \Pr\left(|A_l \hat{v}^*| > (d_1 + d_2\sqrt{2\log n})\epsilon|_{F=f,\hat{v}^*=v}\right)$$

And this proves the lemma. $\square$

*Remark* 4.2. Note that core of the proof is based on the upper bound for $\|\hat{v}^*\|_2$ in Equation 28, which is established by an upper bound for $\|w\|_1$. Therefore, if we relax the assumption $\|w\|_\infty \leq \epsilon$ to the $\ell_1$ assumption $\|w\|_1 \leq n\epsilon$, this lemma still holds true. However, as suggested by Equation 23, we still need a constant $\ell_\infty$ bound on $w$ (say $\|w\|_\infty \leq \frac{1}{3}$) to ensure correct recovery after thresholding. This observation will be used in Section 5.

Part **A** of Theorem 3.4 now follows by combining Equation 23 and Lemma 4.9 and picking $\epsilon = \frac{1}{2}\left(1 + d_1 + d_2\sqrt{\log n}\right)^{-1}$.

### 4.3 General Sparsity Case: Proof of Theorems 3.4(B)

The bounding techniques in the last section has to be modified for the general sparsity case $k \leq \alpha n$. In this section, we describe the extension to the general case. Note that the bounds developed here for the general sparsity case also provide bounds for the linear regime. However, the $SNR$ requirements here are slightly more conservative. Indeed as our statement of Theorem 3.4(B) suggests we need additional constraints on the noise $w$ to ensure perfect support detection with high probability.

The main reason why linear sparsity result of Lemma 4.9 does not generalize to the case when $k$ is small can be attributed to two reasons:

- Significantly higher *effective* noise power: Note that when we only assume that $\|w\|_\infty \leq 1$ and the sparsity of $x$ satisfies $k \ll n$ the ratio of signal power to noise power can be vanishingly small, i.e., as small as $O(k/n)$. For the linear regime this ratio scales as $\Omega(1)$.

- Near singularity of the null space matrix $A$: Note that when $k$ is sublinear with respect to $n$, the matrix $A$ is nearly square and the result of Lemma 4.6 no longer applies.

Recall the setup of Part (**B**):

$$y = G(x + w), \ \|w\|_\infty \leq \epsilon, \ \|w\|_1 \leq \frac{k\epsilon}{\sqrt{\log n}} \tag{31}$$

We point out that in this representation we have absorbed the noise level into the noise $w$, which is different from the situation considered in Eq. 15 for the linear regime.

The main focus in this section is to establish that for $\epsilon \leq 1/\sqrt{\log(n)}$ support recovery is ensured. Our steps mirror those required for the linear regime. Specifically, we can define $\gamma_0$ equivalently as in Equation 17 adapted to the setting of Eq. 31 above.

Again we appeal to the $\ell_2$ bound of Theorem 3.1. This result holds for all sparsity levels. Given the above setup, the solution to the problem of Equation 22 satisfies:

$$\|A\hat{v}\|_2 \leq C_s \frac{\|w\|_1}{\sqrt{k}} \leq C_s \sqrt{k}\epsilon/\sqrt{\log n}$$

with high probability.

Before proceeding to find bounds for $\varepsilon$ we need to deal with the issue of non-singularity of $A$. We quote a recent result of Rudelson [23]:

**Theorem 4.10** (Rudelson). *Let $X$ be an $N \times n$ matrix whose entries are i.i.d. standard Gaussian $\mathcal{N}(0,1)$. Denote $\delta = (N-n)/n$ and $\sigma_{\min}$ as the smallest singular value of $X$. Then for any $t$ such that $\bar{C}n^{-3/2} < t < \bar{c}\delta$,*
$$\Pr\left(\sigma_{\min} \leq t\delta \cdot \sqrt{n}\right) \leq \hat{C}\exp(-cn) + (t/\bar{c}\delta)^{\delta n}$$
*where $\hat{C}, \bar{C}, c, \bar{c}$ are constants.*

Adapting Theorem 4.10 to our context, we have

**Lemma 4.11.** *Let $A$ be an $n \times (n-m)$ matrix whose entries are i.i.d. Gaussian $\mathcal{N}(0, \frac{1}{n-m})$. Then for any $t$ such that $\bar{C}(n-m)^{-3/2} < t < \bar{c}m/(n-m)$,*
$$\Pr\left(\sigma_{\min}(A) \leq \left(\frac{m}{n-m}\right)^2 \cdot \frac{\bar{c}}{2e}\right) \leq \hat{C}\exp(-c(n-m)) + \exp(-\bar{c}m/e). \tag{32}$$

*Proof.* Denote $X = \sqrt{n-m}A$ and we apply Theorem 4.10 with $\delta = m/(n-m)$ and $t = \frac{m}{n-m} \cdot \frac{\bar{c}}{e}$,
$$\Pr\left(\sigma_{\min}(A) \leq \left(\frac{m}{n-m}\right)^2 \sqrt{\frac{n}{n-m}} \cdot \frac{\bar{c}}{2e}\right) \leq \hat{C}\exp(-c(n-m)) + \exp(-\bar{c}m/e). \tag{33}$$

Note that $\frac{n}{n-m} > 1$ and the above equation can be simplified to
$$\Pr\left(\sigma_{\min}(A) \leq \left(\frac{m}{n-m}\right)^2 \cdot \frac{\bar{c}}{2e}\right) \leq \hat{C}\exp(-c(n-m)) + \exp(-\bar{c}m/e). \tag{34}$$
□

Parallel to the Definition 4.1 and Equation 27, we define the typical set $\mathcal{A}_1$ for the general case.

**Definition 4.2.** *Assume $\hat{v}$ is the optimal solution of the optimization problem $\min_v \|x+w+Av\|_1$. We denote $\mathcal{A}_1$ as the intersection of two sets:*
$$\mathcal{A}_1 = \left\{A : \|A\hat{v}\|_2 \leq C_s \cdot \frac{\|w\|_1}{\sqrt{k}}\right\} \cap \left\{A : \sigma_{\min}(A) \geq \left(\frac{m}{n-m}\right)^2 \frac{\bar{c}}{2e}\right\} \tag{35}$$
*and the auxiliary variable $\hat{v}^c$ is defined as*
$$\hat{v}^c = \arg\min_{v:\|v\|_2 \leq \epsilon n^2} \left\{Fv + \sum_{i=k+1}^n |(w+Av)_i|\right\} \tag{36}$$
*where $F = \sum_{i=1}^k sgn(x_i)A_i$.*

**Remark 4.3.** *The superscript $c$ in $\hat{v}^c$ stands for "**c**onstrained".*

Now we have the following property for $\mathcal{A}_1$ and $\hat{v}^c$.

**Lemma 4.12.** *Assume $\epsilon \leq \gamma_0$ in Equation 31. Then we have*

1. *If $A \in \mathcal{A}_1$, then $\hat{v} = \hat{v}^c$;*
2. $\Pr(A \in \mathcal{A}_1) \geq 1 - e^{-c_1 m} - \hat{C}e^{-c(n-m)} - e^{-\bar{c}m/e}$;

3. Suppose $m = \tilde{C}k\log(n/k)$ and $\min\{c_1\tilde{C}, \bar{c}\tilde{C}/e\} \geq 6$, and $A_i$ is the ith row of $A$, then,

$$\mathbb{E}[(A_i\hat{v}^c)^2] \leq \frac{\epsilon^2 C_s^2}{\log n}(1 + o(1)), \quad \forall i = 1, \cdots, k. \tag{37}$$

*Proof.* We know from Lemma 4.4 that if $A \in \mathcal{A}_1$, $\hat{v}$ is the solution of

$$\min_v \left\{ Fv + \sum_{i=k+1}^n |(w + Av)_i| \right\}.$$

Furthermore, if $A \in \mathcal{A}_1$ it also has lower-bounded smallest singular value. This implies

$$\|\hat{v}\|_2 \leq \sigma_{\min}^{-1}(A)\|A\hat{v}\|_2 \leq \epsilon C_s \sqrt{\frac{k}{\log n}} \left(\frac{n-m}{m}\right)^2 \frac{2e}{\bar{c}} \leq \epsilon n^2 \tag{38}$$

where the last inequality holds true when $m = \Theta(k\log(n/k))$. Equation 38 implies that $\hat{v}$ is actually located in the feasible region of minimization problem 36. Therefore part (1) of the lemma follows.

We know from Lemma 4.7 that $\Pr\{A : \|A\hat{v}\|_2 \leq C_s \cdot \frac{\|w\|_1}{\sqrt{k}}\} \geq 1 - e^{-c_1 m}$. Combining this result with Equation 32, part (2) of the lemma follows.

Now we want to prove part (3). First, we have the following bound due to Theorem 3.1:

$$\frac{1}{k}\sum_{i=1}^k (A_i\hat{v}^c)^2 \leq \frac{C_s^2\|w\|_1^2}{k^2} \leq \epsilon^2 C_s^2 \frac{1}{\log n}, \quad \text{if } A \in \mathcal{A}_1$$

Define the partition $\{\mathcal{B}_i\}_i$ of $\mathcal{A}_1^c$: $\mathcal{B}_1 = \mathcal{A}_1^c \cap \{A : \|A\|_F^2 \leq 2n\}$ and $\mathcal{B}_i = \mathcal{A}_1^c \cap \{A : i \cdot n \leq \|A\|_F^2 < (i+1)n\}$ when $i \geq 2$. Here $\|A\|_F = \left(\sum_{i,j} A_{ij}^2\right)^{1/2}$ denotes the Frobenius norm of matrix $A$.

When $A \in \mathcal{B}_1$, we have a loose bound

$$\frac{1}{k}\sum_{i=1}^k (A_i\hat{v}^c)^2 \leq \frac{1}{k}\|A\hat{v}^c\|_2^2 \leq \frac{1}{k}\|A\|_F^2\|\hat{v}^c\|_2^2 \leq \frac{1}{k}(2n)(\epsilon n^2)^2 < \epsilon^2 n^5$$

where the second last inequality follows from the definition of $\mathcal{B}_1$ and $\hat{v}^c$ lies in the feasible set $\{v : \|v\|_2 \leq \epsilon n^2\}$.

When $A \in \mathcal{B}_i (i = 2, 3, \cdots)$, we have

$$\frac{1}{k}\sum_{i=1}^k (A_i\hat{v}^c)^2 \leq \frac{1}{k}\|A\hat{v}^c\|_2^2 \leq \frac{1}{k}\|A\|_F^2\|\hat{v}^c\|_2^2 \leq \frac{1}{k}(i+1)n(\epsilon n^2)^2 < \epsilon^2 n^5 i$$

On the other hand, the probability measures of $\mathcal{B}_i$ satisfies,

$$\Pr(\mathcal{B}_1) \leq \Pr(\mathcal{A}_1^c) \leq e^{-c_1 m} + \hat{C}e^{-c(n-m)} + e^{-\bar{c}m/e}$$

$$\Pr(\mathcal{B}_i) \leq \Pr\{A : i \cdot n \leq \|A\|_F^2\} \leq \left(\frac{e^{(i-1)/2}}{\sqrt{i}}\right)^{-n(n-m)}$$

where the last inequality follows from the tail probability of $\chi^2$ distribution (note that $\|(n-m)A\|_F^2$ is $\chi^2$ distributed with degree $n(n-m)$). It is easy to check that $\frac{e^{(i-1)/2}}{\sqrt{i}} \geq \sqrt{\frac{2}{3}i}$ if $i \geq 2$. Hence

$$\Pr(\mathcal{B}_i) \leq \left(\frac{2}{3}i\right)^{-n^2/4}, \quad \forall i \geq 2$$

Then we can bound $\mathbb{E}[(A_i \hat{v}^c)^2]$ as follows.

$$\begin{aligned}
\mathbb{E}[(A_i \hat{v}^c)^2] &= \mathbb{E}\left[\frac{1}{k}\sum_{i=1}^{k}(A_i \hat{v}^c)^2\right] \\
&= \mathbb{E}\left[\frac{1}{k}\sum_{i=1}^{k}(A_i \hat{v}^c)^2 \Big| A \in \mathcal{A}_1\right]\Pr(\mathcal{A}_1) + \mathbb{E}\left[\frac{1}{k}\sum_{i=1}^{k}(A_i \hat{v}^c)^2 \Big| A \in \mathcal{B}_1\right]\Pr(\mathcal{B}_1) \\
&\quad + \sum_{i=2}^{\infty}\mathbb{E}\left[\frac{1}{k}\sum_{i=1}^{k}(A_i \hat{v}^c)^2 \Big| A \in \mathcal{B}_i\right]\Pr(\mathcal{B}_i) \\
&\leq \epsilon^2 C_s^2 \frac{1}{\log n} + \epsilon^2 n^5 \left(e^{-c_1 m} + \hat{C} e^{-c(n-m)} + e^{-\bar{c}m/e}\right) + \sum_{i=2}^{\infty}\epsilon^2 n^5 i \left(\frac{2}{3}i\right)^{-n^2/4}
\end{aligned}$$

When $\min\{c_1 \tilde{C}, \bar{c}\tilde{C}/e\} \geq 6$, it is easy to check that

$$\lim_{n\to\infty} n^5 \left(e^{-c_1 m} + \hat{C} e^{-c(n-m)} + e^{-\bar{c}m/e}\right)\log n \leq \lim_{n\to\infty} n^5 \left(e^{-6\log n} + e^{-6\log n}\right)\log n = \lim_{n\to\infty}\frac{2\log n}{n} = 0.$$

We can also show $\sum_{i=2}^{\infty} n^5 i \left(\frac{2}{3}i\right)^{-n^2/4} \log n \xrightarrow{n\to\infty} 0$:

$$\begin{aligned}
\sum_{i=2}^{\infty} n^5 i \left(\frac{2}{3}i\right)^{-n^2/4} \log n &= n^5 \left(\frac{2}{3}\right)^{-n^2/4} \sum_{i=2}^{\infty} i^{-n^2/4+1} \log n \\
&\leq n^5 (2/3)^{-n^2/4} \log n \int_{i=2}^{\infty} x^{-n^2/4+1} \mathrm{d}x \\
&= n^5 (2/3)^{-n^2/4} 2^{-n^2/4+2}(n^2/4 - 2)^{-1} \log n \\
&\leq 16 n^3 (4/3)^{-n^2/4} \log n
\end{aligned}$$

$\square$

*Remark 4.4.* It is easy to check numerically that when $n$ is reasonably large (e.g., $n \geq 20$), the $o(1)$ term in equation 37 is actually smaller than one. Therefore, in the later discussion we assume $n \geq 20$ and

$$\mathbb{E}[(A_i \hat{v}^c)^2] \leq \frac{2\epsilon^2 C_s^2}{\log n}, \quad \forall i = 1, \cdots, k. \tag{39}$$

**Lemma 4.13.** *Assume $\epsilon \leq \gamma_0$ in Equation 31 so that support detection is guaranteed. We have*

$$\Pr\left(\max_{i \in \{1,\cdots,k\}} |A_i \hat{v}| > \frac{\sqrt{2}\epsilon C_s}{(\log n)^{1/4}} + 2\epsilon C_s (\log n)^{1/4}\right) \leq e^{-c_1 m} + \hat{C} e^{-c(n-m)} + e^{-\bar{c}m/e} + \frac{2}{\sqrt{\log n}}.$$

*Proof.* Denote $S_F = \mathbb{E}[(A_i \hat{v}^c)^2 | F]$. Note that by the symmetry of $A_i$, all $\mathbb{E}[(A_i \hat{v}^c)^2 | F]$'s should take the same value and therefore $S_F$ does not depend on $i$.

From equation 39, we have $\mathbb{E}_F[S_F] \leq \frac{2\epsilon^2 C_s^2}{\log n}$. Define the set $\mathcal{F}_0 = \{F : S_F > \frac{2\epsilon^2 C_s^2}{\sqrt{\log n}}\}$. Then we know that $\mathcal{F}_0$ has negligible probability measure:

$$\begin{aligned}
\Pr(\mathcal{F}_0) &\leq \left(\Pr(\mathcal{F}_0)\mathbb{E}[S_F | F \in \mathcal{F}_0] + \Pr(\mathcal{F}_0^c)\mathbb{E}[S_F | F \in \mathcal{F}_0^c]\right)\left(\mathbb{E}[S_F | F \in \mathcal{F}_0]\right)^{-1} \\
&= \mathbb{E}_F[S_F]\left(\mathbb{E}[S_F | F \in \mathcal{F}_0]\right)^{-1} \leq \frac{1}{\sqrt{\log n}}
\end{aligned}$$

On the other hand, conditioned on $F$, $A_i(i=1,\cdots,k)$ and $\hat{v}^c$ are independent (c.f. equation 36). Therefore, we can regard $A_i\hat{v}^c|_F$ as a Gaussian random variable and we use $\mu_F$ and $\sigma_F^2$ to denote its mean and variance. Now we consider the case $F \notin \mathcal{F}_0$.

From the above discussion, we know that in this case $\mu_F^2 + \sigma_F^2 \leq \frac{2\epsilon^2 C_s^2}{\sqrt{\log n}}$. This implies that $\mu_F \leq \frac{\sqrt{2}\epsilon C_s}{(\log n)^{1/4}}$ and $\sigma_F^2 \leq \frac{2\epsilon^2 C_s^2}{\sqrt{\log n}}$. From the tail probability of Gaussian distribution, we have

$$\Pr(|A_i\hat{v}^c| > \mu_F + \sigma_F \cdot t | F \notin \mathcal{F}_0) \leq 2Q(t) \leq \frac{2}{t\sqrt{2\pi}} e^{-t^2/2}$$

Substitute $t$ with $\sqrt{2\log n}$ and substitute $\mu_F$ and $\sigma_F$ with the above bound, we have

$$\Pr(|A_i\hat{v}^c| > \frac{\sqrt{2}\epsilon C_s}{(\log n)^{1/4}} + 2\epsilon C_s(\log n)^{1/4} | F \notin \mathcal{F}_0) \leq \frac{1}{n\sqrt{\pi \log n}}$$

Finally we apply the union bound and have

$$\Pr(\max_{i\in\{1,\cdots,k\}} |A_i\hat{v}^c| > \frac{\sqrt{2}\epsilon C_s}{(\log n)^{1/4}} + 2\epsilon C_s(\log n)^{1/4} | F \notin \mathcal{F}_0) \leq \frac{k}{n\sqrt{\pi \log n}}. \tag{40}$$

Furthermore, we know from Lemma 4.12 that if $A \in \mathcal{A}_1$, $\hat{v} = \hat{v}^c$. For simplicity of notation we denote our objective as

$$\mathcal{E} := \left\{ \max_{i\in\{1,\cdots,k\}} |A_i\hat{v}| > \frac{\sqrt{2}\epsilon C_s}{(\log n)^{1/4}} + 2\epsilon C_s(\log n)^{1/4} \right\}$$

and correspondingly

$$\mathcal{E}^c := \left\{ \max_{i\in\{1,\cdots,k\}} |A_i\hat{v}^c| > \frac{\sqrt{2}\epsilon C_s}{(\log n)^{1/4}} + 2\epsilon C_s(\log n)^{1/4} \right\}.$$

Under this notation, Equation 40 can simplified to

$$\Pr(\mathcal{E}|F \in \mathcal{F}_0^c) \leq \frac{k}{n\sqrt{\pi \log n}}$$

Hence we can bound the probability of the unconditioned event $\mathcal{E}$.

$$\begin{aligned}
\Pr(\mathcal{E}) &= \Pr(\mathcal{E}|F \in \mathcal{F}_0^c \text{ and } A \in \mathcal{A}_1) \Pr(\mathcal{F}_0^c \cap \mathcal{A}_1) + \Pr(\mathcal{E}|F \in \mathcal{F}_0 \text{ or } A \in \mathcal{A}_1^c) \Pr(\mathcal{F}_0 \cup \mathcal{A}_1^c) \\
&= \Pr(\mathcal{E}^c|F \in \mathcal{F}_0^c \text{ and } A \in \mathcal{A}_1) \Pr(\mathcal{F}_0^c \cap \mathcal{A}_1) + \Pr(\mathcal{E}|F \in \mathcal{F}_0 \text{ or } A \in \mathcal{A}_1^c) \Pr(\mathcal{F}_0 \cup \mathcal{A}_1^c) \\
&\leq \Pr(\mathcal{E}^c|F \in \mathcal{F}_0^c \text{ and } A \in \mathcal{A}_1) + \Pr(\mathcal{F}_0 \cup \mathcal{A}_1^c)
\end{aligned}$$

The second term is already bounded by $1/\sqrt{\log n} + e^{-c_1 m} + \hat{C}e^{-c(n-m)} + e^{-\bar{c}m/e}$. And the first term can also be bounded,

$$\begin{aligned}
\Pr(\mathcal{E}^c|F \in \mathcal{F}_0^c \text{ and } A \in \mathcal{A}_1) &\leq (\Pr(\mathcal{E}^c|F \in \mathcal{F}_0^c \text{ and } A \in \mathcal{A}_1) \Pr(\mathcal{A}_1) \\
&\quad + \Pr(\mathcal{E}^c|F \in \mathcal{F}_0^c \text{ and } A \in \mathcal{A}_1^c) \Pr(\mathcal{A}_1^c))(\Pr(\mathcal{A}_1))^{-1} \\
&= \Pr(\mathcal{E}^c|F \in \mathcal{F}_0^c)(\Pr(\mathcal{A}_1))^{-1} \\
&\leq \frac{k}{n\sqrt{\pi \log n}} \left(1 - e^{-c_1 m} - \hat{C}e^{-c(n-m)} - e^{-\bar{c}m/e}\right)^{-1} \leq \frac{1}{\sqrt{\log n}}
\end{aligned}$$

$\square$

We are now ready to establish the proof of Theorem 3.4(**B**). From Lemma 4.13 it follows that for $\epsilon \leq \log(n)^{-1/2}/5C_s$ the worst-case perturbation is smaller than $1/2$ with high probability. Decomposing as in Equation 23, and applying Lemma 4.1 it follows that the support detection is guaranteed. This is the statement of Theorem 3.4(**B**).

# 5 Output Noise Model

In this section, we present the results for the output noise model (Equation 1) by converting it to an equivalent input noise model (Equation 2).

The goal of this section is to prove a parallel result to Theorem 3.4 and 3.5.

**Theorem 5.1.** *Consider the setup of Eq. 1. We fix the number of measurements to $3m$, where, $m \geq c_2 \log\left(\frac{n}{2k}\right) 2k$ (see Eq. 13, which arises from the RIP property). We now consider two separate cases:*

**(A)** *Linear sparsity, namely, $\alpha n/2 \leq k \leq \alpha n$ for some $\alpha > 0$ such that the RIP constant $\delta_{2\alpha n} \leq 1/7$ is satisfied. For this case we fix the SNR to satisfy $SNR \geq \tau_A \log n$ for some constant $\tau_A$; Then the TBP+OLS algorithm achieves zero false positives and negatives with high probability, namely,*

$$\Pr(N_m = 0, N_f = 0) > 1 - \frac{4}{\sqrt{\pi \log n}} - 2.24^{-m} - 2.24^{-(n-m)} - 2e^{-c_1 m} - 4e^{-\frac{(\sqrt{n}-\sqrt{n-m})^2}{8}}.$$

**(B)** *General sparsity ($k \leq \alpha n$ with $\alpha$ as in **(A)**). For this case we fix the SNR to satisfy $SNR \geq \tau_B \log^3 n$ for some constant $\tau_B$. Then the TBP+OLS algorithm achieves zero false positives and negatives with high probability, namely,*

$$\Pr(N_m = 0, N_f = 0) > 1 - e^{-c_1 m} - \hat{C} e^{-c(n-m)} - e^{-\bar{c}m/e} - 2.24^{-m} - \frac{5}{\sqrt{\log n}} - 2e^{-\frac{(\sqrt{n}-\sqrt{m})^2}{8}}.$$

*Remark* 5.1. The proof of this theorem is based on the link between the input and the output noise model. We first present the "essential" equivalence between the two models and then point out the modifications needed in adapting the proof of Theorem 3.4 to this proof.

*Remark* 5.2. Compared to Theorem 3.4 and 3.5, an extra $\log n$ factor is required for the SNR level in part **(B)** of the theorem. This $\log n$ factor arises from the looseness of the general case bounds for input noise.

To prove the theorem, we consider the following equation for $w$:

$$Gw = e.$$

This is an over-determined equation with infinitely many possible solutions for $w$. Our approach is to choose the *minimum norm* solution [24] for $w$, namely, $w = G^T(GG^T)^{-1}e$.

Next we establish that this solution results in a satisfactory choice. Suppose the singular value decomposition (SVD) of $G^T = U\Sigma V^T$ where $U \in \mathbb{R}^{n \times m}$ and $\Sigma, V \in \mathbb{R}^{m \times m}$, then we have

$$w = U\Sigma^{-1}V^T e. \tag{41}$$

To see this note that $(GG^T)^{-1} = V\Sigma^{-2}U^T$. Then it follows that $G^T(GG^T)^{-1} = U\Sigma^{-1}V^T$. To express the relation between $w$ and $e$ quantitatively via Equation 41, we need the following two lemmas.

**Lemma 5.2.** *Suppose $e$ is independent Gaussian noise with distribution $\mathcal{N}(0, \epsilon_i^2)$ and $U \in \mathbb{R}^{n \times m}$ is an orthonormal matrix. Denote $\epsilon = \max_i \epsilon_i$. Then*

$$\|Ue\|_\infty \leq \epsilon\sqrt{2 \log n} \quad \text{with probability} \geq 1 - \frac{1}{\sqrt{\pi \log n}}.$$

*Proof.* Let $U_i$ be the ith row of $U$. Then we know that $U_i e$ is still a Gaussian variable with zero mean and variance $\leq \epsilon^2$. Hence, from the union bound and the tail probability of Gaussian distribution,

$$\mathbf{P}(\|Ue\|_\infty > t\epsilon) \leq \sum_{i=1}^n \mathbf{P}(|U_i e| > t\epsilon) \leq \frac{2n}{\sqrt{2\pi}} \cdot \frac{e^{-t^2/2}}{t} \tag{42}$$

Taking $t = \sqrt{2 \log n}$ in the above inequality, we have,

$$\|Ue\|_\infty \leq \epsilon \sqrt{2 \log n} \quad \text{with probability} \geq 1 - \frac{1}{\sqrt{\pi \log n}}$$

□

The next lemma is a classical result on the concentration property of the smallest and largest singular values of Gaussian matrix $G$ (see [22] for example).

**Lemma 5.3** ([22]). *Suppose $G \in \mathbb{R}^{m \times n}$ is a random matrix such that each entry $G_{ij} \sim \mathcal{N}(0, \frac{1}{m})$. We also assume $(n, m)$ satisfies $n = Cm$ where $C > 1$. Then the smallest singular value $\sigma_{\min}$ and the largest singular value $\sigma_{\max}$ of $G$ satisfies the following inequality:*

$$\mathbb{P}(\sqrt{C} - 1 - t/\sqrt{m} \leq \sigma_{\min} \leq \sigma_{\max} \leq \sqrt{C} + 1 + t/\sqrt{m}) \geq 1 - 2e^{-t^2/2}$$

Combining the above two lemmas, we have the following bound for $\|w\|_\infty$ and $\|w\|_1$.

**Lemma 5.4.** *Suppose $Gw = e$ where $G_{ij}$ is i.i.d Gaussian $\mathcal{N}(0, \frac{1}{m})$ and $n = Cm$ and $C > 1$. If $e \sim \mathcal{N}(0, \epsilon^2 I^{m \times m})$, then the minimum norm solution $w = G^T(GG^T)^{-1}e$ satisfies*

$$\|w\|_\infty \leq \frac{2\epsilon}{\sqrt{C} - 1} \sqrt{2 \log n}$$

*with probability $\geq 1 - \frac{1}{\sqrt{\pi \log n}} - e^{-\frac{(\sqrt{n}-\sqrt{m})^2}{8}}$, and*

$$\|w\|_1 \leq 2\sqrt{2C}(\sqrt{C} - 1)^{-1} m\epsilon$$

*with probability $\geq 1 - 2.24^{-m} - e^{-\frac{(\sqrt{n}-\sqrt{m})^2}{8}}$.*

*Proof.* See Appendix. □

For Theorem 5.1(**A**) in the linear regime we note that $\epsilon = 1/\sqrt{\tau_A \log n}$. It follows that, we have an equivalent input noise model from Lemma 5.4 with probability $\geq 1 - \frac{1}{\sqrt{\pi \log n}} - 2.24^{-m} - 2e^{-\frac{(\sqrt{n}-\sqrt{m})^2}{8}}$ such that

$$y = G(x + w), \|w\|_\infty \leq \frac{2\sqrt{2}}{(\sqrt{C} - 1)\sqrt{\tau_A}}, \|w\|_1 \leq 2\sqrt{2C^{-1}}(\sqrt{C} - 1)^{-1} \cdot \frac{n}{\sqrt{\tau_A \log n}}.$$

According to the explanation in Remark 4.2, these two assumptions on $w$ can replace the original conditions in Theorem 3.4(**A**) and still ensure correct support detection if $\tau_A$ is sufficiently large.

For Theorem 5.1(**B**) the variance of the ith component of noise $e_i$ is $\epsilon^2 = \frac{1}{\tau_B \log^3 n}$ for some sufficiently large constant $\tau_B$. It follows from Lemma 5.4 that the output noise model $y = Gx + e$ is essentially equivalent to following input noise model,

$$y = G(x + w), \|w\|_\infty \leq \frac{2\sqrt{2}}{(\sqrt{C} - 1)\sqrt{\tau_B \log n}}, \|w\|_1 \leq 2\sqrt{2C}(\sqrt{C} - 1)^{-1} \cdot \frac{m}{\sqrt{\tau_B \log n}}.$$

with probability $\geq 1 - \frac{1}{\sqrt{\pi \log n}} - 2.24^{-m} - 2e^{-\frac{(\sqrt{n}-\sqrt{m})^2}{8}}$. They match the conditions in Theorem 3.4(**B**) if $\tau_B$ is sufficiently large.

Hence we can regard the output model as the related input noise model and mimic the steps of the proof in the last section. The main steps remain unchanged except for the perturbation computation in Lemma 4.9. The difficulty is that when we solve $Gw = e$ from the minimum norm criterion and get $w = G^T(GG^T)^{-1}e$, $w$ is weakly correlated with $G$ and hence bounding techniques developed above might fail to work. This problem can be handled in the following way.

The fundamental step of the proof to Lemma 4.9 is to bound the inner product of $A_l$ and $\hat{v}$. To accomplish this, we used the fact that $\hat{v}$ depends on $A_l$ only through $F$. This type of reasoning can also be extended to the model $y = Gx + e$. First, we choose the minimum norm $w = G^T(GG^T)^{-1}e$ and we have $y = G(x + w)$. Now, $\hat{v}$ depends on $A_l$ not only through $F$ but also through $w$ because $w = G^T(GG^T)^{-1}e$ might potentially depend on $A_l$. Therefore in the next step of bounding $A_l\hat{v}$, we need to condition on both $w$ and $F$ to ensure that $A_l$ and $\hat{v}$ are conditionally independent of each other. The minimum norm $w$ has the property that $w$ is in the range space of $G^T$, which implies $A^T w = 0$. Alternatively from a QR decomposition of $G^T = QR$ we see that $w$ can be represented as $w = QR^{-T}e$. It is well known that $Q$ and $R$ are independent if $G$ is originally a Gaussian matrix (c.f. [25]). Therefore if we suppose $R^{-T}e$ to be fixed (but unknown) then no information about $Q$ can be deduced from $w$ besides $A^T w = 0$ (i.e., $w \in \text{span}(Q)$). Furthermore $w$ can be assumed to be uniformly distributed on a sphere for the purpose of analysis. Particularly, this implies that the conditional distribution $p(A_l|w) = p(A_l|A^T w = 0)$. Next, the conditional distribution $A_l|_{A^T w = 0}$ is still Gaussian and the knowledge of $w$ only reveals average value of the rows of $A$, which is similar to dependency of $A_l$ through $F$ we had in Lemma 4.9. Consequently, identical steps can be followed to establish the main result as well. Finally similar arguments as those used in Section 7.2 show that an OLS step will remove all the false alarms.

## 6 Numerical Examples

Our first example illustrates the performance difference between LASSO and basis pursuit(i.e., only Step 1 of TBP). In this example, we choose the signal dimension $x$ to be 200 and set 10% of components to be nonzero. The sensing matrix $G$ we use here is a $100 \times 200$ matrix, each element of which is i.i.d. Gaussian. Without loss of any generality we let the nonzero components to be the first $k$ components of the signal. The effective SNR of the system is $6 \log n$. The reconstruction result is shown in Figure 2. From this example, we can see that while LASSO does as well as our algorithm in recovering the support, the amplitude values appear to be biased.

We also recall Figure 1 of Section 1.2 for a more systematic comparison between these two approaches. One main difficulty we found in implementing LASSO was to determine the optimal tuning parameter, $\lambda$. The analysis of [11] suggests that $\lambda = 2\sigma\sqrt{2 \log n}$, where $\sigma$ is the variance of i.i.d additive Gaussian noise, would be good choice. On the other hand [26] recommends $\lambda = 2\sigma\sqrt{n}$. Note that in both these instances we need to know the noise level. In our experimentation we also found that support recovery could be improved when $\lambda$ is allowed to depend on the number of measurements $m$ and the sparsity level $k$ as well. However, this is in general very difficult.

For Fig. 1 (left figure) we varied the sparsity level, $k$, while keeping the $n = 200$, $m = 100$ and $SNR = 6 \log(n)$ fixed. We only implemented the first two steps of TBP, i.e., we ignored the OLS step. To implement LASSO we experimented with different values of $\lambda$ and plotted the best parameter we could find. Specifically, for the left figure we optimized the error probability over $\lambda$ for specific sparsity levels via exhaustive search and we observed that $\lambda = 0.2$ worked best. We fixed this value of $\lambda$ for all sparsity levels. We see that TBP significantly outperforms LASSO. The success rate of LASSO begins to drop around $k = 10$ whereas the success rate of TBP begins to drop around $k = 30$. The phase transition of TBP happens much later compared to LASSO.

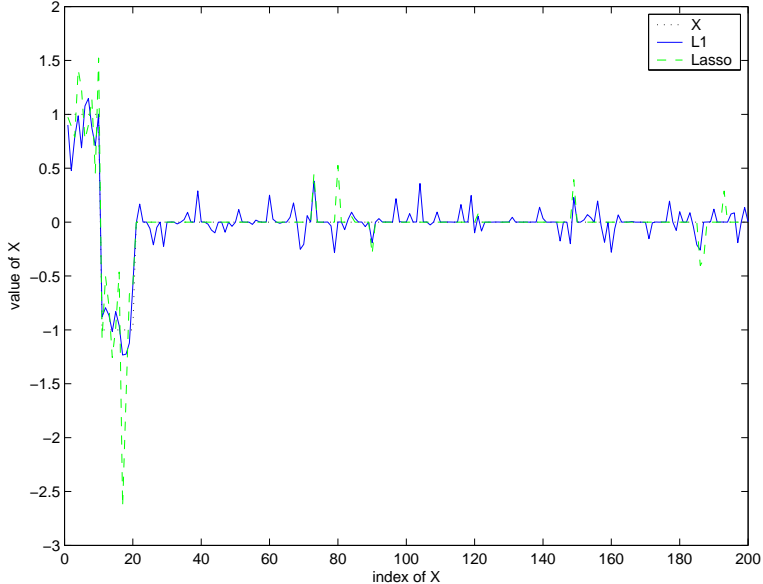

Figure 2: LASSO vs. $\ell_1$ minimization(basis pursuit). The signal length is $m = 200$, 5% of components are 1's, 5% of components are $-1$'s and the rest 90% of components are 0's. $G$ is a random $100 \times 200$ matrix and $SNR = 6\log n$. LASSO is a biased estimator and gives poor reconstruction at nonzero components.

For the second experiment (right figure in Fig. 1) we fixed $n = 200$, $k = 10$ and varied the number of measurements $m$. Since $m > k$ the plot starts at $m = 20$. Each point on the plot corresponds to an average over 80 Monte-Carlo trials. To get a good value for the tuning parameter, $\lambda$, we again looked at specific measurement levels and optimized for the success probability. The optimal $\lambda$ turned out to be around 0.3. This was then fixed varying values of $m$. Again we see that TBP performs much better compared to LASSO.

There are possibly two reasons for the poor performance of LASSO. First, we believe thresholding is really necessary even for shrinkage operator such as LASSO and second, we might not always be able to choose the optimal $\lambda$. We are unaware of any results regarding how $\lambda$ adapts to different $k$ and $m$. As pointed out in the previous discussion, this might be a serious problem in practice.

In the last experiment (Figure 3), we show how SNR level influences the probability of success in TBP. Here we implement only the first two steps of TBP(Basis Pursuit and thresholding) and do not use the extra regression step. Next we fix the signal dimension $n$, sparsity $k$ and number of measurements $m$ and simulate the results for different levels of SNR. Specifically, we fix $(n, m, k) = (200, 100, 20)$ and $x_{\min} = 1$. We let $\sigma^{-1} = (2\sqrt{12\log n} + 2) \cdot \theta$, and vary $\theta$ from $10^{-2}$ to $10^2$ which varies the $SNR$. From our theory, we expect to see the phase transition around $\Theta(\log n)$, which is what is observed here. Each point on the curve(i.e., each SNR level) is an average of 200 Monte Carlo trials.

We can see from Figure 3(a,b) that the success probability curve jumps from zero to one around $\theta = 10^0$ for both the Gaussian and Bernoulli ensembles. Note that while our theory is based on the Gaussian ensemble, it appears that the results are not particularly sensitive to non-Gaussian ensembles. The simulation results also suggests that the OLS step in Section 3.3 (step 3 and 4) is not really necessary. Also in experimentation we did not find any qualitative or quantitative difference between sublinear and linear scenarios. Therefore it is an open problem whether the $SNR$ gap between these two regimes in Theorem 5.1(**B**) can be improved.

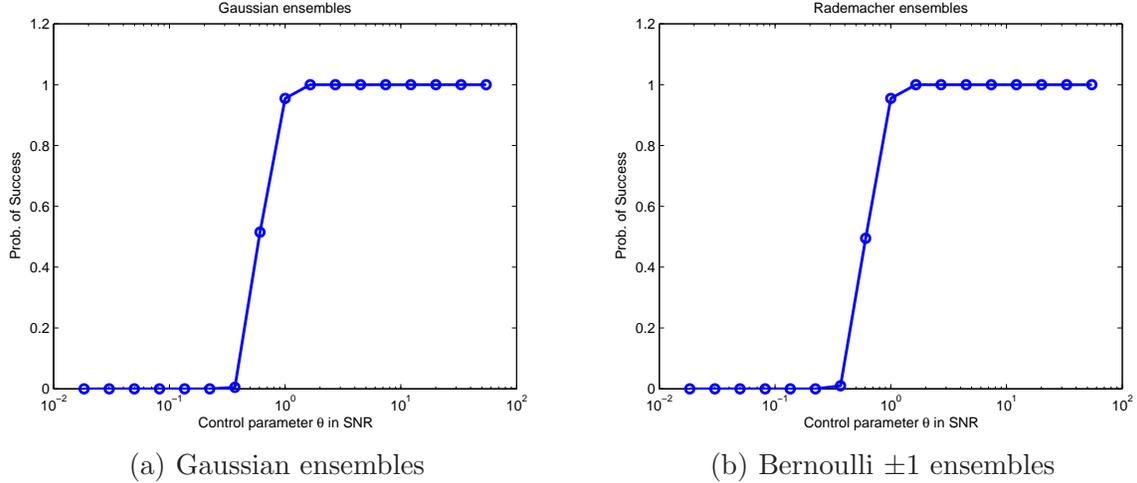

(a) Gaussian ensembles

(b) Bernoulli ±1 ensembles

Figure 3: Success probability of support recovery for TBP as a function of SNR for the output noise model. Here we fix $(n, m, k) = (200, 100, 20)$ and $x_{\min} = 1$. We then let $SNR = \frac{1}{\sigma^2}$ vary. Specifically, we let $\sigma^{-1} = (2\sqrt{12 \log n} + 2) \cdot \theta$, and vary $\theta$ from $10^{-2}$ to $10^2$. Each point on the curve is an average of 200 Monte Carlo trials. The phase transition happens around $\theta = 10^0$. (a) Gaussian ensembles: each component $G_{ij}$ of the sensing matrix is i.i.d. from $\mathcal{N}(0, \frac{1}{m})$; (b) Bernoulli ±1 ensembles: each component $G_{ij}$ is independently chosen to be either $\frac{1}{\sqrt{m}}$ or $-\frac{1}{\sqrt{m}}$ with equal probability.

# 7 Appendix

## 7.1 $\ell_2$ Approximation: Proof of Theorem 3.1

In this section we prove Theorem 3.1 with respect to the sensing model:

$$y = G(x + w)$$

Denote $z := x + w$. We let $T_0$ be the indices of the largest $|T_0| = k$ components of $z = x + w$. We further define the rest indices as $T_1, \cdots, T_J$ of equal size $|T_j| = M, j \geq 1$ (where $M$ is an design parameter and will be specified later), by decreasing order of magnitude. We also use $T_{01}$ to denote $T_0 \cup T_1$.

Denote the reconstruction error $h := \hat{x} - z$ and we have,

$$\begin{aligned} \|z_{T_0}\|_1 - \|h_{T_0}\|_1 - \|z_{T_0^c}\|_1 + \|h_{T_0^c}\|_1 &\leq \|z_{T_0} + h_{T_0}\|_1 + \|z_{T_0^c} + h_{T_0^c}\|_1 \\ &= \|\hat{x}\|_1 \leq \|z\|_1 \end{aligned}$$

which can be simplified to

$$\|h_{T_0^c}\|_1 \leq \|h_{T_0}\|_1 + 2\|z_{T_0^c}\|_1. \tag{43}$$

Next, we relate the $\ell_2$ norm of $h_{T_{01}^c}$ to the $\ell_1$ norm of $h_{T_0}$. It is obvious that the $k$th largest components of $h_{T_0^c}$ satisfies

$$|h_{T_0^c}(k)| \leq \|h_{T_0^c}\|_1/k$$

Squaring both sides and then summing up from $k = M + 1$ up to $k = n$, we have,

$$\|h_{T_{01}^c}\|_2^2 \leq \|h_{T_0^c}\|_1^2 \sum_{k=M+1}^{n} \frac{1}{k^2} \leq \|h_{T_0^c}\|_1^2 \sum_{k=M+1}^{n} \frac{1}{(k-1)k}$$
$$= \|h_{T_0^c}\|_1^2 \sum_{k=M+1}^{n} \left(\frac{1}{k-1} - \frac{1}{k}\right) \leq \frac{\|h_{T_0^c}\|_1^2}{M}. \tag{44}$$

Combining inequality 43 and 44, we have,

$$\|h_{T_{01}^c}\|_2 \leq \frac{\|h_{T_0}\|_1 + 2\|z_{T_0^c}\|_1}{\sqrt{M}} \leq \sqrt{\frac{k}{M}} \left(\|h_{T_0}\|_2 + \frac{2\|z_{T_0^c}\|_1}{\sqrt{k}}\right) \leq \sqrt{\frac{k}{M}} \left(\|h_{T_0}\|_2 + \frac{2\|w\|_1}{\sqrt{k}}\right),$$

where the second inequality follows from the Cauchy-Schwartz inequality.

Hence,

$$\|h\|_2 \leq \|h_{T_{01}}\|_2 + \|h_{T_{01}^c}\|_2 \leq \left(1 + \sqrt{\frac{k}{M}}\right) \|h_{T_{01}}\|_2 + 2\sqrt{\frac{k}{M}} \cdot \frac{\|w\|_1}{\sqrt{k}}. \tag{45}$$

From the above inequality, we can see that the task remaining is to upper bound $\|h_{T_{01}}\|_2$. Before deriving this bound, we first derive a bound for $\sum_{j \geq 2} \|h_{T_j}\|_2$ as an intermediate step.

Observe that the magnitude of each components in $T_{j+1}$ is bounded by the average of the magnitudes in $T_j$:

$$|h_{T_{j+1}}(k)| \leq \|h_{T_j}\|_1 / M.$$

Then by taking squares at both sides and then summing up from $k = jM + 1$ up to $k = (j+1)M$,

$$\|h_{T_{j+1}}\|_2^2 \leq \|h_{T_j}\|_1^2 / M$$

We take the square-root of both sides and sum up from $j = 1$ up to the end:

$$\sum_{j \geq 2} \|h_{T_j}\|_2 \leq \sum_{j \geq 1} \frac{\|h_{T_j}\|_1}{\sqrt{M}} = \frac{\|h_{T_0^c}\|_1}{\sqrt{M}} \tag{46}$$

Combining with inequality 43, we have

$$\sum_{j \geq 2} \|h_{T_j}\|_2 \leq \sqrt{\frac{k}{M}} \left(\|h_{T_0}\|_2 + \frac{2\|w\|_1}{\sqrt{k}}\right) \tag{47}$$

Now $\|h_{T_{01}}\|_2$ can be bounded from the RIP property in the following way.

$$0 = \|Gh\|_2 = \|G_{T_{01}} h_{T_{01}} + \sum_{j \geq 2} G_{T_j} h_{T_j}\|_2 \geq \|G_{T_{01}} h_{T_{01}}\| - \sum_{j \geq 2} \|G_{T_j} h_{T_j}\|_2$$
$$\geq \sqrt{1 - \delta_{M+k}} \|h_{T_{01}}\|_2 - \sqrt{1 + \delta_M} \sum_{j \geq 2} \|h_{T_j}\|_2$$
$$\geq \sqrt{1 - \delta_{M+k}} \|h_{T_{01}}\|_2 - \sqrt{1 + \delta_M} \sqrt{\frac{k}{M}} \left(\|h_{T_0}\|_2 + \frac{2\|w\|_1}{\sqrt{k}}\right)$$
$$\geq \left(\sqrt{1 - \delta_{M+k}} - \sqrt{1 + \delta_M} \sqrt{\frac{k}{M}}\right) \|h_{T_{01}}\|_2 - \sqrt{1 + \delta_M} \sqrt{\frac{k}{M}} \frac{2\|w\|_1}{\sqrt{k}}.$$

where the second last inequality follows from inequality 47. This implies that

$$\|h_{T_{01}}\|_2 \leq \sqrt{1+\delta_M}\sqrt{\frac{k}{M}}\frac{2\|w\|_1}{\sqrt{k}}/C_M, \tag{48}$$

where $C_M = \sqrt{1-\delta_{M+k}} - \sqrt{1+\delta_M}\sqrt{\frac{k}{M}}$. Finally, combining inequality 45 and 48, we got

$$\|h\|_2 \leq \left(\frac{\sqrt{1+\delta_M}(1+\sqrt{\frac{k}{M}})}{C_M} + 1\right)\frac{2\|w\|_1}{\sqrt{M}}. \tag{49}$$

We choose $M = 2k$ such that $C_M$ are positive constants and this proves the theorem.

Note that we need to ensure $C_M$ to be positive, i.e, $\sqrt{1-\delta_{k+2k}} - \sqrt{1+\delta_{2k}}\sqrt{\frac{k}{2k}} > 0$ which implies

$$(1+\delta_{2k})\frac{1}{2} + \delta_{3k} < 1. \tag{50}$$

In [27], the authors prove that for positive integer $c$ and $r$, it follows that $\delta_{cr} \leq c \cdot \delta_{2r}$. Applying this inequality in condition 50, we only need to ensure $\delta_{2k} \leq \frac{1}{7}$.

## 7.2 Proof of Theorem 3.5

We only prove part **(A)** and part **(B)** follows along similar reasoning. First note that the number of false positives $N_f \leq m - k$ since any optimal solution to LP is a basic feasible solution. Suppose $G_{2,I}^\dagger$ is the pseudo-inverse of $G_{2,I}$. Then

$$G_{2,I}^\dagger y_2 = G_{2,I}^\dagger(G_2(x+w_2)) = x + w_{2,I} + G_{2,I}^\dagger G_{2,I^c}w_{2,I^c}. \tag{51}$$

We denote $e_2 := G_{2,I^c}w_{2,I^c}$. Since each element of $G_{2,I^c}$ is i.i.d. Gaussian with variance $\frac{1}{m}$, each component of $e_2$ is also i.i.d. Gaussian with variance

$$\mathrm{Var}(e_2) \leq \frac{n-|I|}{m}\epsilon^2 \leq \frac{n}{m}\epsilon^2 \leq \left(\frac{1}{8+2d_1+2d_2\sqrt{2\log n}} \cdot \frac{1}{2\sqrt{2\log n}}\right)^2 \triangleq \epsilon_1^2$$

The singular value decomposition(SVD) of $G_{2,I}$ gives us $G_{2,I} = U\Sigma V^T$ where $U, V$ are orthonormal matrices and $\Sigma \in \mathbb{R}^{2m \times (k+N_f)}$ is a diagonal matrix. By [24] the pseudo-inverse of $G_{2,I}$ is $G_{2,I}^\dagger = V\Sigma^\dagger U^T$, where $\Sigma^\dagger$ is the pseudo-inverse of $\Sigma$.

Now the reconstruction error can be represented as $G_{2,I}^\dagger e_2 = V\Sigma^\dagger U^T e_2$. Since $e_2$ is i.i.d. Gaussian as shown above and $U$ are orthonormal matrix, $U^T e_2$ is still i.i.d Gaussian with the same distribution as $e_2$. This means $(\Sigma^\dagger U^T n_2)_i$ is independent Gaussian variable with variance $\leq \frac{\epsilon_1^2}{\Sigma_{ii}}$.

The matrix $G_{2,I}$ has $2m$ rows and $k + N_f \leq m$ columns. The smallest singular value of $G_{2,I}$ is $\geq \frac{1}{2}(1-\frac{1}{\sqrt{2}})$. Hence the variance of $(\Sigma^\dagger U^T e_2)_i$ is $\leq \frac{4\epsilon^2}{(1-\frac{1}{\sqrt{2}})^2}$ By applying Lemma 5.2, we have

$$\|V(\Sigma^\dagger U^T e_2)\|_\infty \leq 2\epsilon_1(1-\frac{1}{\sqrt{2}})^{-1}\sqrt{2\log n}$$
$$\leq \frac{1}{(1-\frac{1}{\sqrt{2}})(8+2d_1+2d_2\sqrt{\log n})} \leq \frac{3}{8}, \quad \text{w.p.} \geq 1 - \frac{1}{\sqrt{\pi \log n}}.$$

It is clear from the assumption of the Theorem that $\|w_{2,I}\|_\infty \leq \frac{1}{8}$ and finally we can bound the reconstruction error in equation 51 as

$$\|w_{2,I} + G_{2,I}^\dagger G_{2,I^c}w_{2,I^c}\|_\infty \leq \|w_{2,I}\|_\infty + \|G_{2,I}^\dagger G_{2,I^c}w_{2,I^c}\|_\infty \leq \frac{1}{8} + \frac{3}{8} = \frac{1}{2}$$

## 7.3 Proof of lemma 5.4

Suppose $e \sim \mathcal{N}(0, \epsilon^2 I^{m \times m})$. We write the SVD of $G^T$ as $G^T = U\Sigma V^T$ where $U \in \mathbb{R}^{n \times m}, V \in \mathbb{R}^{m \times m}$ are orthonormal matrices and $\Sigma \in \mathbb{R}^{m \times m}$ are diagonal matrix. Then $w$ can be reformulated as

$$w = U\Sigma^{-1}V^T e \tag{52}$$

Since $V$ is orthonormal, $V^T e$ is still Gaussian with the same distribution as $e \sim \mathcal{N}(0, \epsilon^2 I^{m \times m})$.

Conditioned on all $\Sigma_{ii}$'s being lower-bounded by $\frac{1}{2}(\sqrt{C} - 1)$, the variance of $(\Sigma^{-1}V^T e)_i$ is $\leq \frac{4\epsilon^2}{(\sqrt{C}-1)^2}$. By applying Lemma 5.2, by conditioning on $G$ we have

$$\|w\|_\infty \leq 2\epsilon(\sqrt{C} - 1)^{-1}\sqrt{2 \log n} \quad \text{with probability} \geq 1 - \frac{1}{\sqrt{\pi \log n}}.$$

On the other hand, the concentration property of smallest singular value in Lemma 5.3 implies that $\Sigma_{ii} > \frac{1}{2}(\sqrt{C} - 1)$ with probability $\geq 1 - e^{-\frac{(\sqrt{n}-\sqrt{m})^2}{8}}$. Therefore by applying Lemma 5.2

$$\|w\|_\infty \leq 2\epsilon(\sqrt{C} - 1)^{-1}\sqrt{2 \log n} \quad \text{with probability} \geq 1 - \frac{1}{\sqrt{\pi \log n}} - e^{-\frac{(\sqrt{n}-\sqrt{m})^2}{8}}.$$

To compute the $\ell_1$ bound we proceed as follows: we bound the squared $\ell_2$ for a fixed $G$. Note that for a fixed $G$, the noise $w$ is a zero mean Gaussian random variable as before. We know that all the singular values of $G$ are lower bounded by $(\sqrt{C} - 1)/2$ with probability $\geq 1 - e^{-\frac{(\sqrt{n}-\sqrt{m})^2}{8}}$. The following computation are done for a given $G$ whose smallest singular values of $G$ are lower bounded by $(\sqrt{C} - 1)/2$.

From Cauchy-Schwarz inequality, we have $\|w\|_1^2 \leq n\|w\|_2^2$. We know from the previous discussion that

$$\|w\|_2^2 = \sum_{i=1}^{m}(\Sigma^{-1}V^T e)_i^2$$

where each $(\Sigma^{-1}V^T e)_i$ are independent zero-mean Gaussian r.v with variance upper bounded by $\frac{4\epsilon^2}{(\sqrt{C}-1)^2}$.

Suppose $t$ is $\chi^2$ random variable with degree $m$. We have

$$\Pr\left(\|w\|_2^2 \leq \frac{4\epsilon^2}{(\sqrt{C} - 1)^2} \cdot 2m \Big| G\right) \geq \Pr(t \leq 2m) \geq 1 - 2.24^{-m} \tag{53}$$

where the last inequality follows from the tail probability of $\chi^2$ distribution.

Finally if we take into account of all possible $G$'s, we have

$$\Pr\left(\|w\|_1^2 \leq \frac{4\epsilon^2}{(\sqrt{C} - 1)^2} \cdot 2mn\right) \geq 1 - 2.24^{-m} - e^{-\frac{(\sqrt{n}-\sqrt{m})^2}{8}} \tag{54}$$

The lemma follows by noting that $n = Cm$.

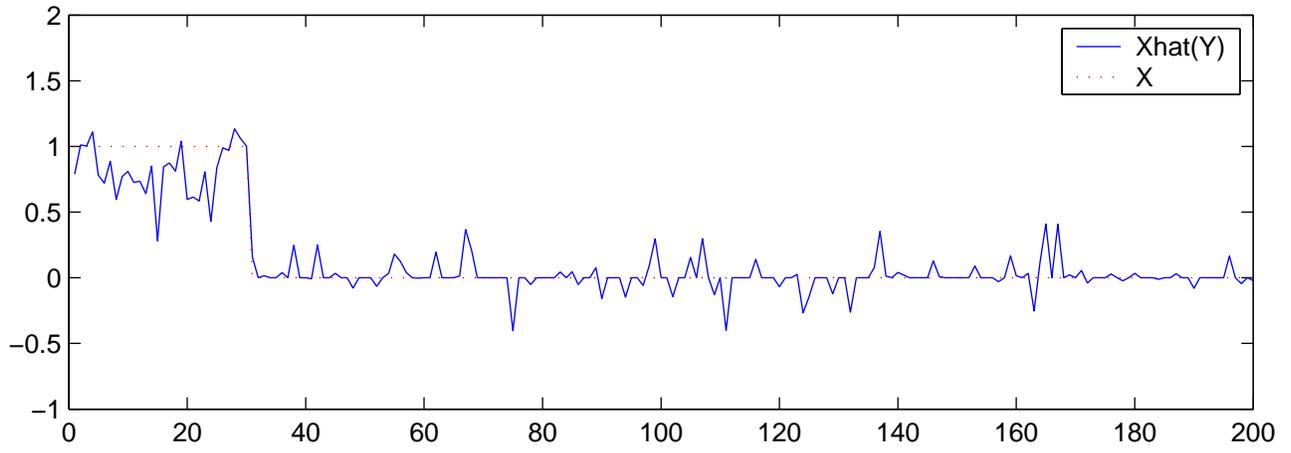
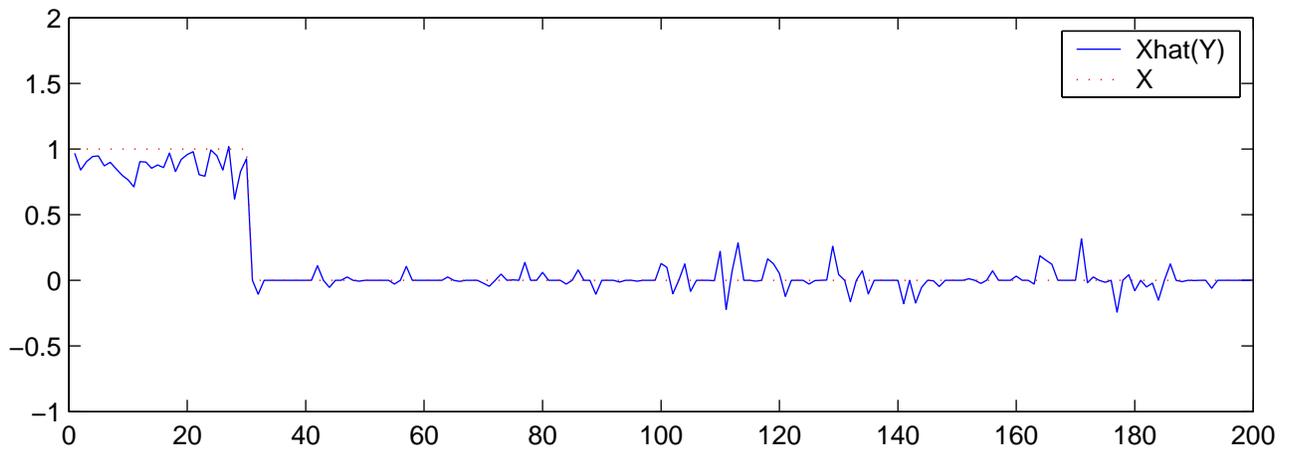

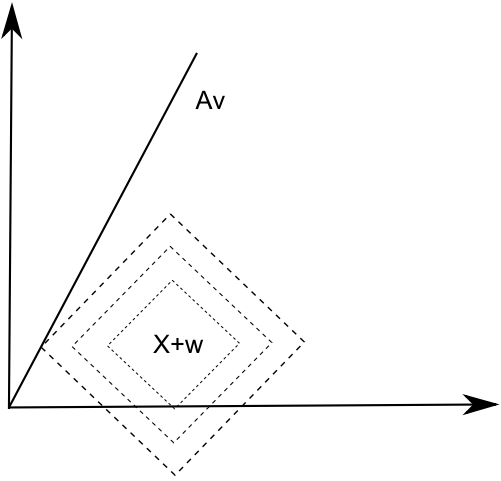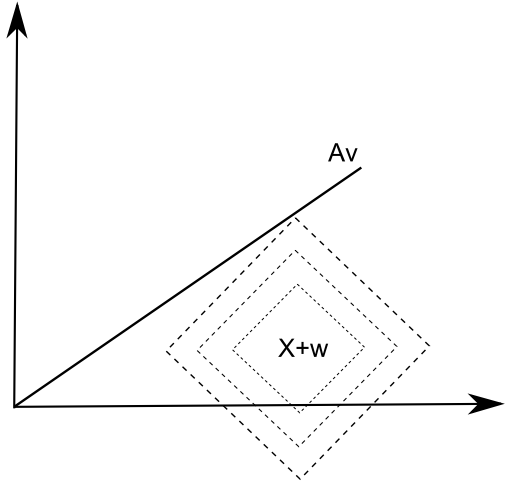

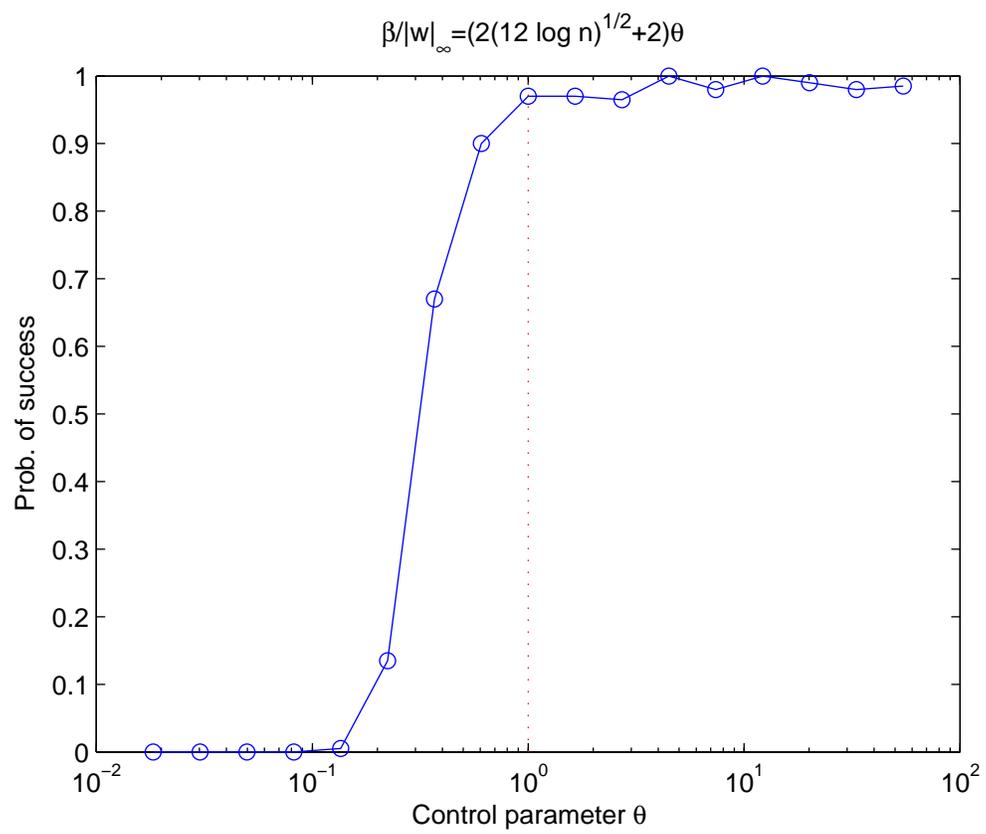